\documentclass[longauth]{aa}  
\usepackage{graphicx}
\usepackage[]{hyperref}
\usepackage{xcolor}
\usepackage{float}
\usepackage{subcaption}
\usepackage{makecell}
\usepackage{multirow}
\hypersetup{
  colorlinks   = true, 
  urlcolor     = blue, 
  linkcolor    = blue, 
  citecolor   = blue 
}

\newcommand{\Swift}{\textit{Swift}}
\newcommand{\SN}{SN~2025kzr}

\usepackage{txfonts}
\begin{document} 

   \title{The flash-ionised SN~Ibn~2025kzr: H-free CSM formed during a precursor outburst 55 days prior to explosion}

   \author{S. de Wet \inst{1} \corrauth{sndwe@dtu.dk}
   \and G. Leloudas\inst{1}\email{giorgos@space.dtu.dk}
   \and D.~A.~H. Buckley\inst{2,3,4} \email{dibnob@saao.ac.za}
   \and N. Erasmus\inst{2,5}\email{nerasmus@saao.ac.za}
   \and P.~J. Groot\inst{2,3,6,7}\email{p.groot@astro.ru.nl}
   \and E.~A. Zimmerman\inst{8}\email{erez.zimmerman@weizmann.ac.il}
   \and P. Chen\inst{9,10}\email{chenp1220@gmail.com}
   \and Y. Tampo\inst{2,3}\email{yusuke@saao.ac.za}
   \and M. Pursiainen\inst{11}\email{miika.pursiainen@warwick.ac.uk}
   \and T. Killestein\inst{11}\email{tom.killestein@gmail.com}
   \and F. Stoppa\inst{12}\email{fiorenzo@blackgem.org}
   \and G.~K. Jaisawal\inst{1}\email{gaurava@space.dtu.dk}
   \and A. Gal-Yam\inst{8}\email{avishay.gal-yam@weizmann.ac.il}
   \and K. Maeda\inst{13}\email{keiichi.maeda@kusastro.kyoto-u.ac.jp}
   \and J. Anderson\inst{14}\email{janderso@eso.org}
   \and T.-W. Chen\inst{15}\email{twchen@gm.astro.ncu.edu.tw}
   \and M. Gromadzki\inst{16}\email{marg@astrouw.edu.pl}
   \and C.~P. Guti\'errez\inst{17,18}\email{cgutierrez@ice.csic.es}
   \and E. Kankare\inst{19}\email{erkki.kankare@utu.fi}
   \and T.~E. M\"uller-Bravo\inst{20,21}\email{t.e.muller-bravo@tcd.ie}
   \and T. Pessi\inst{14}\email{thallis.pessi@mail.udp.cl}
   \and S. Smartt\inst{12,22}\email{stephen.smartt@physics.ox.ac.uk}
   \and J. Sollerman\inst{23}\email{jesper@astro.su.se}
   \and L. Tartaglia\inst{24}\email{tleonardo.tartaglia@gmail.com}
   \and D.~R. Young\inst{22}\email{david.young.rse@gmail.com}
   \and M.~R. Alarcon\inst{25,26,27}\email{miguel.rodriguez@iac.es}
   \and K.~W. Smith\inst{12,22}\email{K.W.Smith@qub.ac.uk}
   \and H.~F. Stevance\inst{12,22}\email{hfstevance@gmail.com}
   \and T. de Boer\inst{28}\email{tdeboer@hawaii.edu}
   \and K. Chambers\inst{28}\email{chambers@ifa.hawaii.edu}
   \and C.-C. Lin\inst{28}\email{cclin33@hawaii.edu}
   \and T.~B. Lowe\inst{28}\email{tlowe@hawaii.edu}
   \and P. Minguez\inst{28}\email{pdl31@hawaii.edu}
   \and M. Nicholl\inst{22}\email{matt.nicholl@qub.ac.uk}
   \and G.~S.~H. Paek\inst{28}\email{gpaek@hawaii.edu}
   \and R.~J. Wainscoat\inst{28}\email{rjw@hawaii.edu}
   }

\institute{
DTU Space, Technical University of Denmark, Elektrovej 327--328, DK-2800 Lyngby, Denmark
\and South African Astronomical Observatory, PO Box 9, 7935 Observatory, South Africa
\and Department of Astronomy, University of Cape Town, Private Bag X3, Rondebosch 7701, South Africa
\and Department of Physics, University of the Free State, PO Box 339, Bloemfontein 9300, South Africa
\and Department of Physics, Stellenbosch University, Stellenbosch 7600, South Africa
\and Department of Astrophysics/IMAPP, Radboud University, PO Box 9010, 6500 GL Nijmegen, The Netherlands
\and The Inter-University Institute for Data Intensive Astronomy, University of Cape Town, Private Bag X3, Rondebosch 7701, South Africa
\and Department of Particle Physics and Astrophysics, Weizmann Institute of Science, 234 Herzl St, 7610001 Rehovot, Israel
\and Institute for Advanced Study in Physics, Zhejiang University, Hangzhou 310027, China
\and Institute for Astronomy, School of Physics, Zhejiang University, Hangzhou 310027, China
\and Department of Physics, University of Warwick, Gibbet Hill Road, Coventry CV4 7AL, UK
\and Department of Physics, University of Oxford, Denys Wilkinson Building, Keble Road, Oxford OX1 3RH, UK
\and Department of Astronomy, Kyoto University, Kitashirakawa-Oiwake-cho, Sakyo-ku, Kyoto 606-8502, Japan
\and European Southern Observatory, Alonso de C\'ordova 3107, Vitacura, Casilla 19001, Santiago, Chile
\and Graduate Institute of Astronomy, National Central University, 300 Jhongda Road, 32001 Jhongli, Taiwan
\and Astronomical Observatory, University of Warsaw, Al. Ujazdowskie 4, 00-478 Warszawa, Poland
\and Institute of Space Sciences (ICE, CSIC), Campus UAB, Carrer de Can Magrans, s/n, E-08193 Barcelona, Spain
\and Institut d'Estudis Espacials de Catalunya (IEEC), Edifici RDIT, Campus UPC, 08860 Castelldefels (Barcelona), Spain
\and Department of Physics and Astronomy, University of Turku, 20014 Turku, Finland
\and School of Physics, Trinity College Dublin, The University of Dublin, Dublin 2, Ireland
\and Instituto de Ciencias Exactas y Naturales (ICEN), Universidad Arturo Prat, Chile
\and Astrophysics Research Centre, School of Mathematics and Physics, Queen's University Belfast, Belfast BT7 1NN, UK
\and Department of Astronomy, Oskar Klein Center, Stockholm University, SE-106 91 Stockholm, Sweden
\and INAF -- Osservatorio Astronomico d'Abruzzo, via Mentore Maggini snc, I-64100 Teramo, Italy
\and Light Bridges, Observatorio Astron\'omico del Teide, Carretera del Observatorio del Teide, s/n, E-38570 G\"u\'imar, Tenerife, Spain
\and Instituto de Astrof\'isica de Canarias (IAC), C/ V\'ia L\'actea, s/n, E-38205 La Laguna, Tenerife, Spain
\and Departamento de Astrof\'isica, Universidad de La Laguna (ULL), E-38206 La Laguna, Tenerife, Spain
\and Institute for Astronomy, University of Hawaii, 2680 Woodlawn Drive, Honolulu HI 96822}

   \date{Latest draft: \today}

  \abstract{
Type Ibn supernovae (SNe) are a class of interacting SNe characterised by narrow helium lines in their spectra. We present an extensive observational dataset of the Type Ibn \SN{} at 51 Mpc, including the discovery of a precursor outburst with a peak brightness of $M_r{\approx}-13.6$~mag beginning ${\approx}55$ days before explosion. Our photometry indicates the SN was discovered within the first day of explosion, and showing fast-rising, ultraviolet-bright emission peaking at $M_r=-19.26\pm0.09$~mag and a peak blackbody temperature of $T{\approx}29000$~K, consistent with shock breakout within a region of dense and confined circumstellar material (CSM). Our high-cadence spectroscopic dataset spanning 1.9--58.5 days post-explosion shows flash-ionised emission features during the first 10 days. In our SALT spectrum at 3.8 days we observe a pronounced blueshift of the \ion{He}{ii} lines by 460~km~s$^{-1}$ compared to the \ion{He}{i} lines at zero velocity, while a Pickering-decrement analysis reveals a CSM that is fully hydrogen-free. The timing of the disappearance of the flash features combined with the CSM velocity of 1500~km~s$^{-1}$ imply a mass-loss event ${\approx}66$ days before explosion, in close agreement with the timing of the precursor observed 55 days before explosion and strongly suggestive of a physical link. We derive a CSM mass of 0.03--1.7~$M_\odot$ and a corresponding high mass-loss rate ${\gtrsim}10^{-1}~M_\odot$~yr$^{-1}$. The precursor timescale and energetics suggest an extreme mass-loss event that might be explained by wave-driven mass loss during the late stages of nuclear burning, in particular the oxygen-burning phase. Overall, we favour a single massive Wolf-Rayet progenitor with $M_\mathrm{ZAMS}{\sim}30$--40~$M_\odot$ to explain \SN{}, although a binary origin cannot be excluded.

 }


   \maketitle

\section{Introduction} \label{sec:intro}

Massive stars with zero-age main-sequence masses ($M_\mathrm{ZAMS}$) of ${\gtrsim}8~M_{\odot}$ end their lives as core-collapse supernovae (SNe) following nuclear fuel exhaustion and the subsequent collapse of their stellar cores. Observational evidence indicates that stars with initial masses of ${\sim}8$--20~$M_\odot$ explode as red supergiants (RSGs) and produce Type IIP SNe \citep{Smartt2009}, but the fates of more massive stars (${>}25~M_\odot$) remains unsettled. These stars are expected to undergo significant stripping of their hydrogen (H) envelopes through line-driven stellar winds or binary interaction \citep{Heger2003,Eldridge2008}, forming Wolf-Rayet (WR) stars that explode as SNe of Types IIb, Ib, or Ic: so-called `stripped-envelope' SNe \citep[SESNe;][]{Filippenko1997,Galyam2017HSN,Modjaz2019}. 

In recent years this simple picture has been challenged by the fact that no WR star has yet been identified in pre-explosion imaging in connection with a SN Ib/c \citep[e.g.][]{Eldridge2016,Zhao2025}, while the occurrence rate of SNe Ib/c is too high to be consistent with the observed WR population \citep{Smith2011}. As a consequence, there is a growing consensus that the main pathway to SESNe involves binary systems of lower-mass massive stars which undergo mass transfer \citep{Podsiadlowski1992,Eldridge2008,Eldridge2013,Chen2024}. 

A rare subclass of SESNe are the Type Ibn events characterised by narrow helium (He) emission lines in their spectra \citep{Foley2007,Pastorello2008}. The velocities inferred from the widths of the He lines (around ${\sim}1000$~km~s$^{-1}$) are broadly consistent with the wind velocities of Galactic WR stars \citep{Crowther2007}, suggesting a direct link to WR progenitors. SNe Ibn are among the most luminous classes of SNe, with typical peak absolute magnitudes ranging from $-18$ to $-20$. Their fast rise to peak brightness followed by a rapid decay has led to them being grouped among the rapidly evolving transients \citep{Drout2014,Pursiainen2018,Ho2023}, and their light curves are thought to be powered by the interaction of the SN ejecta with H-poor circumstellar material (CSM). Light curve modelling of a sample of SNe Ibn showed that the steep light curve decay is consistent with a steep CSM density profile ($\rho\propto r^{-3}$), high mass-loss rates of 0.025--0.25~$M_\odot$~yr$^{-1}$ (assuming a wind velocity of 1000~km~s$^{-1}$), and ejecta properties consistent with canonical SESNe \citep{Maeda2022}. The low \textsuperscript{56}Ni masses inferred for SNe Ibn \citep[compared to SESNe and SNe II;][]{Anderson2019,Meza2020} have been interpreted as evidence for high initial masses (${\gtrsim}18~M_\odot$) since fallback of a large amount of ejecta onto the compact object remnant would result in little to no \textsuperscript{56}Ni being ejected \citep{Maeda2022}. 

The H-rich analogues of SNe Ibn are the well-studied SNe IIn \citep{Schlegel1990}, which show significant diversity in their light curves in both duration and luminosity \citep{Kiewe2012,Taddia2013,Nyholm2020,Hiramatsu2024}. This diversity is interpreted as reflecting a wide range in CSM densities and extents. By contrast, SNe Ibn light curves form a rather homogeneous sample \citep{Hosseinzadeh2017} that has been attributed to the dense CSM being confined to smaller radii ($\lesssim10^{15}$~cm) than in SNe IIn \citep{Moriya2016}. 

This confined CSM can be revealed through the detection of recombination emission lines of high-ionisation species that disappear within days of explosion \citep[`flash spectroscopy';][]{Galyam2014,Khazov2016,Yaron2017}. Such features were detected as far back as four decades ago \citep[SN 1983K;][]{Niemela1985}, with famous early examples being SNe~1993J, 1998S, and 2006bp \citep{Benetti1994,Garnavich1994,Fassia2001,Chugai2001,Quimby2007}. The detection of these features requires rapid spectroscopic follow-up within days of explosion, so early cases were mostly limited to fortuitous detections. The advent of wide-field transient surveys led to an increase in the number of detections and the first systematic studies of samples of such events \citep{Khazov2016,Bruch2021,Bruch2023,JacobsonGalan2024}. A surprising finding from these campaigns was that a large fraction (${\gtrsim}1/3$) of Type II SNe show transient emission lines if observed early enough \citep[${\lesssim}2$ days after explosion;][]{Bruch2021}, implying that enhanced mass loss in the months to years before explosion is more widespread than previously thought. In the majority of these cases the objects have been H-rich and have evolved into SNe IIb, IIP, or IIL. By contrast, H-poor SNe with flash features are far less common: no classical SN Ib/c has shown such features, and the H-poor flash-spectroscopy sample is dominated by SNe Ibn, although with only a handful of detections \citep[e.g. SNe~2010al, 2019uo, 2019wep, and others][]{Pastorello2015,Gangopadhyay2020,Gangopadhyay2022}. 

There is a growing consensus that eruptive and binary-driven mass loss predominate over stellar winds in stripping massive stars of their hydrogen envelopes \citep[for a review see][]{Smith2014}. There are several reasons for this: the mass-loss rates from stellar winds of hot massive stars and red supergiants have been revised downwards \citep{Puls2008,Sundqvist2019,Beasor2020}; there is evidence that canonical SESN progenitors are the outcome of the envelope stripping of lower-mass massive stars in binaries \citep{Podsiadlowski1992,Sana2012,Eldridge2013,Lyman2016,Fang2019}; and there is a growing number of transients showing evidence for eruptive mass loss events \citep{Smith2006,Ofek2013,Mauerhan2013}. Observations of SN precursor emission strongly indicate that the outbursts themselves are responsible for creating the dense CSM \citep{Pastorello2008,Ofek2013,Strotjohann2021}. One of the most important questions concerning SNe Ibn and other SNe with dense CSM is identifying the physical mechanism responsible for this mass loss. The timescales of these outbursts (months to years prior to explosion) suggest a connection to the late phases of nuclear burning. Several ideas have been proposed, ranging from wave-driven mass loss to super-Eddington accretion onto a compact object binary companion \citep{Shiode2014,Fuller2017,Fuller2018,Leung2021,Wu2022,Tsuna2024}. 

Among SNe IIn, precursor emission is fairly common, with 25\% showing precursor emission within three months of explosion \citep{Strotjohann2021}. For SNe Ibn, however, only three cases are known \citep[SNe 2006jc, 2019uo, and 2023fyq;][]{Pastorello2008,Strotjohann2021,Brennan2024}. It is likely that this is simply due to an observational bias caused by the rarity of SNe Ibn: they account for ${\sim}1\%$ of all CCSNe in magnitude-limited surveys, whereas SNe IIn are ten times more abundant \citep{Perley2020}. In the volume-limited ATLAS100 survey SNe~Ibn are ${\lesssim}1\%$ the CCSN rate and six times less abundant than SNe IIn \citep{Srivastav2026}. Among the three SNe Ibn with precursors, the rising pre-SN light curve of SN~2023fyq has been interpreted as strong evidence for binary-driven mass loss \citep{Dong2024}, whereas SNe 2006jc and 2019uo show eruptive-type precursor behaviour. Further well-studied examples are needed to elucidate the physical mechanisms responsible for such pre-SN mass loss and determine the progenitor channels for SNe Ibn.  

Here we present the early discovery and follow-up observations of the Type Ibn SN~2025kzr at 51~Mpc. ATLAS, Pan-STARRS and TZF survey data reveal a precursor outburst at ${\sim}55$~days before explosion, making \SN{} only the fourth SN Ibn to show precursor emission. We show here that this precursor lasted at least 30 days. Our high-cadence spectroscopic dataset consisting of 15 spectra obtained within the first 10 days---while flash features were visible---constitutes the most comprehensive dataset of a flash-ionised SN Ibn to date, allowing us to perform a detailed spectroscopic analysis during the flash phase. Taken together, the CSM probed through our flash-spectra suggests a direct link with mass ejected during the precursor eruption. The paper is structured as follows. In Section \ref{sec:observations} we provide details on our observations, and in Section \ref{sec:host} we present the host galaxy and reddening properties of \SN{}. In Sections \ref{sec:photometric_properties} through \ref{sec:X_radio} we investigate the photometric, spectroscopic, polarimetric, X-ray, and radio properties of \SN{}. In Section \ref{sec:discussion} we discuss the notable aspects of \SN{} and develop a picture of the progenitor system. All times are in the observer frame, and all magnitudes are in the AB system unless stated otherwise.

\section{Observations}\label{sec:observations}
\subsection{Discovery}
\SN{} was discovered by the Asteroid Terrestrial-impact Last Alert System \citep[ATLAS;][]{Tonry2018,Smith2020} on 2025 May 22 at 01:05:22~UTC with a $c$-band magnitude of $17.00\pm0.03$ \citep{2025TNSAN.153....1S}. The most recent non-detection occurred three days prior on May 19 at 19:47:31 UTC. The ATLAS Virtual Research Assistant \citep[VRA;][]{Stevance2025,Stevance2026} scored \SN{} highly (${>}8.5$, for scores ranging from 0 to 10) which triggered fully-automated imaging and spectroscopic observations with the Mookodi low-resolution spectrograph and imager \citep{Erasmus2024} mounted on the Lesedi 1~m telescope \citep{Worters2016} at the South African Astronomical Observatory (SAAO) site in Sutherland, South Africa. Our first 600~s spectrum obtained 0.81 days after the ATLAS discovery showed strong flash-ionised emission lines of \ion{C}{iii} and \ion{He}{ii} superposed over a blue continuum \citep{2025TNSAN.154....1W}. We did not report a classification, however, as the SN sub-type was still unclear. About a week after discovery, \SN{} was classified as a Type IIn SN due to its spectroscopic similarity with SN~1998S \citep{IIn_classification}. We will show in this paper that the photometric and spectroscopic properties of \SN{} match those of SNe Ibn, and we therefore regard this as the more appropriate classification. At the time of discovery, \citet{2025TNSAN.153....1S} noted the proximity of the SN \citep[within the 100 Mpc volume of the ATLAS100 survey;][]{Srivastav2026}, the recent upper limits, and the existence of a precursor in Pan-STARRS survey data. This prompted us to instigate an intensive SAAO follow-up campaign, supported by other observatories. 

\subsection{Optical photometry}\label{subsec:phot}
Following the initial classification we scheduled nightly photometry with Mookodi mounted on the SAAO 1~m Lesedi telescope. Observations consisted of 60~s exposures in the Sloan Digital Sky Survey (SDSS) $griz$ bands and 300~s exposures in the $u$-band while the SN was bright. At later times after the SN had faded we increased our exposure time to 300~s in the $griz$ bands and discontinued our $u$-band observations due to its lower sensitivity. Gaps in our light curve coverage were due to inclement weather conditions at Sutherland or scheduling constraints. We computed the SN photometry using an adapted version of the MeerLICHT and BlackGEM pipelines that also performs difference imaging \citep[Vreeswijk et al., in preparation;][]{Groot2024}. For our reference images we obtained deep 600~s exposures more than six months after discovery. The pipeline performs photometric calibration using a dedicated Mookodi calibration catalogue. 

We submitted an urgent request 1.4 days after the ATLAS discovery to obtain \Swift{} Ultra-Violet Optical Telescope \citep[UVOT;][]{Roming2005} target-of-opportunity (ToO) observations of \SN{} in the ultra-violet ($uvw2$, $uvm2$, $uvw1$, $U$) and optical ($B$, $V$) bands. A total of 28 epochs---designated by separate Obs IDs---were obtained over the course of \Swift{} follow-up. We downloaded all publicly available UVOT photometry from the Swift Archive Download portal hosted on the UK Swift Science Data Centre (UKSSDC) website\footnote{\href{https://www.swift.ac.uk/swift_portal/}{https://www.swift.ac.uk/swift\_portal/}} and used the HEASoft Swift FTOOLS software package, version 6.34\footnote{\href{https://heasarc.gsfc.nasa.gov/docs/software/heasoft/}{https://heasarc.gsfc.nasa.gov/docs/software/heasoft/}} to perform all reductions. The very last epoch was specifically requested by us to obtain a deep template of the host galaxy. At later epochs once the SN had faded, we coadded exposures using \verb|uvotimsum| to obtain deeper images. We used the \verb|uvotsource| tool to perform aperture photometry on the images, with the source and background apertures set to $5^{\prime\prime}$ and $10^{\prime\prime}$ in radius, respectively. From these count rates we subtracted the final epoch count rates to remove any host contribution. We report $3\sigma$ limiting magnitudes in cases where the signal-to-noise ratio (SNR) of the host-subtracted count rate is ${<}3$.

We obtained archival difference-imaging forced photometry at the SN position from ATLAS and the Zwicky Transient Facility \citep[ZTF;][]{Bellm2019} via the ATLAS Forced Photometry server\footnote{\href{https://fallingstar-data.com/forcedphot/}{https://fallingstar-data.com/forcedphot/}} \citep{Shingles2021} and the ZTF forced-photometry service \citep{Masci2019}, respectively. We queried all available data for both surveys. The ATLAS data spanned 9.5 years prior to explosion. The fifth ATLAS unit on Tenerife \citep[ATLAS-TDO;][]{Licandro2023} was in operation at the time and obtained unfiltered observations around the time of the precursor. The blue-sensitive CMOS detector provides a similar wavelength coverage to the ATLAS $c$-band filter. We include these data in addition to the public data, and stacked all nightly ATLAS photometry using the routine of \citet{Young_plot_atlas_fp}. We set a 5$\sigma$ and 3$\sigma$ detection threshold for the ATLAS and ZTF photometry, respectively. For ZTF we filtered the data following recommended quality control practices, and compute upper limits at the 5$\sigma$ level, as recommended by \citet{Masci2023}. ATLAS upper limits are also calculated at the 5$\sigma$ level. We additionally obtained Pan-STARRS \citep{Chambers2016} forced photometry at the SN position in the $w$, $i$, and $y$ bands, from the full history Pan-STARRS observations of this field spanning 9.2 years. This included the ongoing use of the Pan-STARRS Near-Earth Object survey data for extragalactic transients \citep{Fulton2025}. Setting a detection threshold of 3$\sigma$, we identify only four detections, all in the $w$ band on MJD 60763. We calculate 3$\sigma$ upper limits for all non-detections. 

We could not continue our monitoring beyond ${\sim}70$ days post-explosion due to solar conjunction. All detections for \SN{} and its precursor are available at the CDS (see the Data availability section). 

\subsection{Optical spectroscopy}\label{subsec:spectra}
In addition to our photometric follow-up, we also scheduled nightly spectroscopic observations with Mookodi. Mookodi has a stepped-slit configuration with narrow ($2^{\prime\prime}$) and wide ($4^{\prime\prime}$) slit widths corresponding to spectral resolutions of $R{\sim}350$ and $R{\sim}175$ at $6000~\AA$, respectively \citep{Erasmus2024}. Due to Mookodi's fully-robotic operation \citep{IO}, the wider slit is usually used for observations of transients to ensure that the majority of the flux is captured. We used the wide slit (mk-wide) for all observations except for the second and third nights where we used the narrow slit (mk-narrow). The exposure time was mostly set to 600~s although from 20 days onwards we increased this to 900~s. For certain epochs we obtained multiple exposures, so in these cases we summed the spectra to obtain a higher signal-to-noise ratio. Each observation was followed by a 4~s arc observation for wavelength calibration. We reduced the data using a Mookodi-specific pipeline adapted from the open-source, python-based Automated SpectroPhotometric REDuction \citep[ASPIRED;][]{Lam2023} toolkit. The pipeline performs trace fitting, optimal spectral extraction, wavelength calibration and flux calibration using observations of a spectrophotometric standard star. 

We obtained three epochs of spectra with the Southern African Large Telescope \citep[SALT;][]{Buckley2006} equipped with the long-slit Robert Stobie Spectrograph (RSS) through observing program 2024-2-LSP-001 (PI Buckley). We used the PG0900 grating with two different grating angles of $13.25^{\circ}$ and $20.00^{\circ}$ to cover both blue (3500--6600~$\AA$) and red (6000--9000~$\AA$) wavelength ranges. The spectral resolution at the central wavelength for the blue and red grating angle setups was 5.7~$\AA$ ($R{\sim}880$) and 5.5~$\AA$ ($R{\sim}1360$), respectively. SALT provides users with cosmic-ray cleaned, bad-pixel corrected, and wavelength-calibrated 2D spectra as part of their standard data products. We extracted the SN trace using the ASPIRED toolkit and performed flux calibration using observations of the spectrophotometric standard star LTT 4364. For each epoch we merged the blue and red spectra by scaling them until the fluxes matched in the overlapping wavelength range. 

We additionally obtained a single 1800~s observation of \SN{} with the dual-beam, fibre-fed, echelle High Resolution Spectrograph (HRS) on SALT, also through program 2024-2-LSP-001. Our observation employed the Low Resolution mode ($R{\sim}14000$) encompassing the wavelength range 3700--8800~$\AA$. Reduced data products from the HRS MIDAS pipeline \citep{Kniazev2016,Kniazev2017} are provided to all SALT users. The pipeline performs flatfield corrections, order extraction, wavelength calibration and sky fiber subtraction. Even though \SN{} was fairly bright at the time of our observations with $r{\approx14.9}$~mag, the signal-to-noise ratio was not high enough to prevent a strong blaze effect from being introduced into the final spectrum. Despite this, we are still able to use our HRS spectrum to obtain an estimate of the Milky Way (MW) and host galaxy extinction as well as the exact redshift of \SN{} (Section~\ref{sec:host}). 

We obtained six epochs of long-slit spectra with EFOSC2 mounted on the ESO New Technology Telescope (NTT) through the ePESSTO+ collaboration\footnote{\href{https://www.pessto.org}{https://www.pessto.org}}. Observations were obtained using gratings 11 (3345--7470~$\AA$) and 16 (6000--9995~$\AA$) with corresponding resolutions of 13.8~$\AA$ ($R{\sim}390$) and 13.4~$\AA$ ($R{\sim}595$), respectively. We reduced the data using the dedicated PESSTO pipeline\footnote{\href{https://github.com/svalenti/pessto}{https://github.com/svalenti/pessto}} \citep{PESSTO}, which performs bias and flatfield corrections, wavelength calibration, and flux calibration through observations of a spectrophotometric standard star. 

We triggered four epochs of spectroscopic follow-up observations with ALFOSC mounted on the Nordic Optical Telescope (NOT) through proposal 71-021 (PI Leloudas). Our observations employed grism 4 covering the wavelength range 3200--9600~\AA~with a resolution of 16.2~\AA~($R{\sim}360$). We reduced the data using the PyNOT\footnote{\href{https://jkrogager.github.io/pynot/}{https://jkrogager.github.io/pynot/}} pipeline, which performs bias and flatfield corrections, wavelength calibration, spectral extraction, and flux calibration using spectrophotometric standard star observations. The first epoch was obtained using an instrument rotation angle $54^{\circ}$ away from the parallactic angle, which resulted in significant slit losses at blue wavelengths due to the high airmass ($\sec z=2.2$) of the target. This prevented us from reliably flux calibrating the spectrum using a standard star. Instead, we performed flux calibration by multiplying the continuum-normalised spectrum with a 17000~K blackbody---the approximate observed blackbody temperature derived from photometry at this time assuming no extinction. The shape of the resulting spectrum shows good agreement with other spectra taken at similar times. 

We obtained a single epoch of observations using the SpUpNIC spectrograph \citep{Crause2019} mounted on the SAAO 1.9~m telescope. Our observations consisted of $3\times600$~s of exposure with grating 6 ($R{\sim1000}$) that covered the wavelength range 4400--7050~$\AA$. The data were reduced using standard tasks in IRAF, including bias and flat-field corrections and wavelength calibration. We flux-calibrated the spectrum using observations of the standard star LTT 3864.

We also obtained a single 900~s epoch of spectroscopy with the Inamori-Magellan Areal Camera and Spectrograph \citep[IMACS;][]{Dressler2011} mounted on the Magellan Baade telescope, using Grism 300 ($R{\sim}1000$). The IMACS data were reduced with the SIMplified SPECtroscopic\footnote{\href{https://astro.subhashbose.com/software/simspec}{https://astro.subhashbose.com/software/simspec}} (SimSpec) reduction pipeline, which is based on {\tt PyRAF}. SimSpec reduction includes bias and flat-field corrections, cosmic-ray removal, wavelength calibration and relative flux calibration with a spectrophotometric standard star observed on the same night as the science object. 

All spectra used in this work are summarised in Table \ref{tab:obs_log} and will be made publicly available through the
Weizmann Interactive Supernova Data Repository \citep[WISeREP][]{WISEREP} upon publication.  

\subsection{Optical imaging polarimetry}
We obtained two epochs of $V$-band imaging polarimetry with NOT/ALFOSC. A single cycle over the half-wave plate angles $0.0\deg$, $22.5\deg$, $45.0\deg$ and $67.5\deg$ was obtained for both epochs. The data were reduced using a custom pipeline presented in \citet{Pursiainen2025} that extracts ordinary and extraordinary beam photometry with $r=2\times~$FWHM (full-width at half-maximum) following \citet{Pursiainen2023}. We probe the effect of Galactic interstellar polarisation (ISP) in three ways. First, in the ALFOSC field-of-view (FOV) there is a single bright star we can use to evaluate the ISP. The stellar polarisation is consistent between the two epochs, settling on a value of $Q\sim0.0\%$ and $U\sim0.3\%$. In comparison, stars in the Heiles catalogue \citep{Heiles2000} within $5\deg$ show similar or smaller levels at the same polarisation angle. Finally, the maximum level of the MW ISP has empirically been shown to be $9\times E(B-V)$ \citep{Serkowski1975}. Using $E(B-V)=0.062\pm0.002$~mag (Table \ref{tab:SN_gal}), the maximum ISP is $\sim0.55$\%, larger but still consistent with the stellar-based measurements, as expected. As such we adopt the polarisation measured for the star ($Q\sim0.00\pm0.04\%$ and $U\sim0.33\pm0.04\%$) in the ALFOSC FOV as the MW ISP along the line of sight.

\subsection{X-rays}
\SN{} was simultaneously observed by the \Swift{} X-Ray Telescope \citep[XRT;][]{2005SSRv..120..165B} over the course of UVOT follow-up. We downloaded all data from the UKSSDC website and processed the data for each Obs ID using the standard \texttt{xrtpipeline} routine. We additionally used the UKSSDC website to build a combined image from all of the XRT data for a total effective exposure time of 52.9~ks. No significant X-ray emission is detected at the SN position in our single-epoch and combined images. We used the \texttt{XIMAGE sosta} tool to calculate $3\sigma$ upper limits on the count rate at the SN position.

Assuming a Galactic column density of $N_\mathrm{H} = 6 \times 10^{20}$~cm$^{-2}$ \citep{2016A&A...594A.116H}, we converted our combined image count rate upper limits to flux upper limits using \texttt{WebPIMMS}\footnote{\url{https://heasarc.gsfc.nasa.gov/cgi-bin/Tools/w3pimms/w3pimms.pl}} under different spectral assumptions. For a non-thermal power-law spectrum with photon index $\Gamma = 2$, the unabsorbed 0.3--10~keV flux is ${<}1.87 \times 10^{-14}$~erg~s$^{-1}$~cm$^{-2}$; for a thermal plasma (\texttt{apec}) model with $kT = 1$~keV, the flux is ${<}1.17 \times 10^{-14}$~erg~s$^{-1}$~cm$^{-2}$; and for a thermal blackbody model with $kT = 1$~keV, the flux is predicted to be ${<}2.26 \times 10^{-14}$~erg~s$^{-1}$~cm$^{-2}$. Based on these estimates, we adopt an upper limit on the X-ray flux of ${\approx} 2 \times 10^{-14}$~erg~s$^{-1}$~cm$^{-2}$, corresponding to an upper limit on the X-ray luminosity of ${\approx} 6 \times 10^{39}$~erg~s$^{-1}$ at the assumed distance of 51~Mpc. We note that this upper limit does not correspond to the instrinsic luminosity, as any X-ray emission may be subject to internal or CSM absorption.

\subsection{Radio}
We obtained a single epoch of MeerKAT continuum observations in the L-band (1.28~GHz) and S4 band (3~GHz) through director's discretionary time proposal DDT-20250915-SD-01 (PI: de Wet) to search for radio emission associated with SN 2025kzr. Our observations were obtained on 2025 September 20 (122~days post-explosion) and consisted of 1.66~hours of integration time per band. We used J0408$-$6545 as the flux and bandpass calibrator for both bands, while J1051$-$2023 and J1037$-$-2934 were used as the complex gain calibrators for the L and S4 bands, respectively. Inspecting the SARAO Science Data Processor (SDP) primary-beam corrected images, we do not detect a radio source near the position of \SN{}. We compute upper limits as three times the RMS noise ($3\sigma$) in a circular aperture with a radius that is five times the synthesised beam size, centred on the SN position. Our upper limits of 49 and 16~$\mu$Jy in the L and S4 bands translate to luminosity limits of $L_{1.28\,\text{GHz}} < 1.5 \times 10^{26}$\,erg\,s$^{-1}$\,Hz$^{-1}$ 
and $L_{3\,\text{GHz}} < 5.0 \times 10^{25}$\,erg\,s$^{-1}$\,Hz$^{-1}$.

\section{Host galaxy and reddening}\label{sec:host}
\begin{figure}
    \centering
    \includegraphics[width=\linewidth]{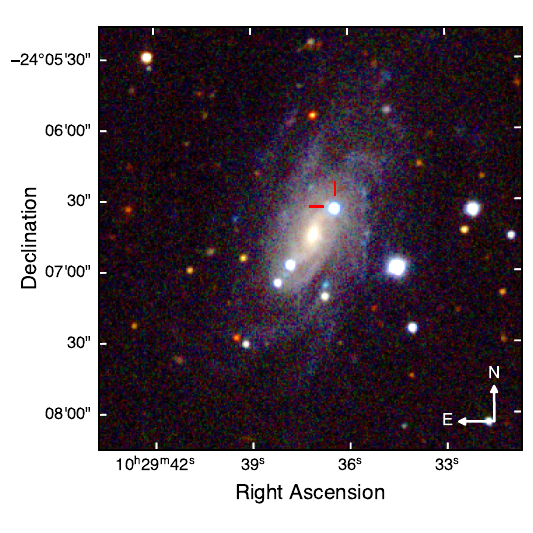}
    \caption{Colour-composite image showing the location of \SN{} within its host galaxy, ESO 501- G 001. The image was created by combining deep, 600~s stack images obtained with Mookodi in the $gri$ bands over the course of our follow-up campaign. }
    \label{fig:host}
\end{figure}

We report the fundamental parameters of \SN{} and its host galaxy in Table \ref{tab:SN_gal}. The coordinates\footnote{\href{https://www.wis-tns.org/object/2025kzr}{https://www.wis-tns.org/object/2025kzr}} of the SN are $\alpha=10^\mathrm{h}29^\mathrm{m}36.43^\mathrm{s}$, $\delta=-24^\circ06^{\prime}33.61^{\prime\prime}$ (J2000), which is offset by $14.8^{\prime\prime}$ from the core of its host galaxy, ESO 501- G 001 (Figure \ref{fig:host}). According to the NASA/IPAC Extragalactic Database\footnote{\href{https://ned.ipac.caltech.edu}{https://ned.ipac.caltech.edu}} (NED) ESO 501- G 001 has an SAB(s)d morphological classification, indicative of a late-type, star-forming galaxy. It is likely that the progenitor of \SN{} occurred in a region of active star formation given its location near an inner spiral arm. The redshift of ESO 501- G 001 is $z=0.012689(7)$. A total of 16 redshift-independent distance modulus estimates are listed for ESO 501- G 001, with the majority of these derived using the Tully-Fisher relation. We compute the weighted mean using the reported uncertainties as weights, and obtain $\mu = 33.54 \pm 0.09$~mag corresponding to a distance of $51.0 \pm 2.0$~Mpc. 

\begin{figure*}
    \centering
    \includegraphics[width=\linewidth]{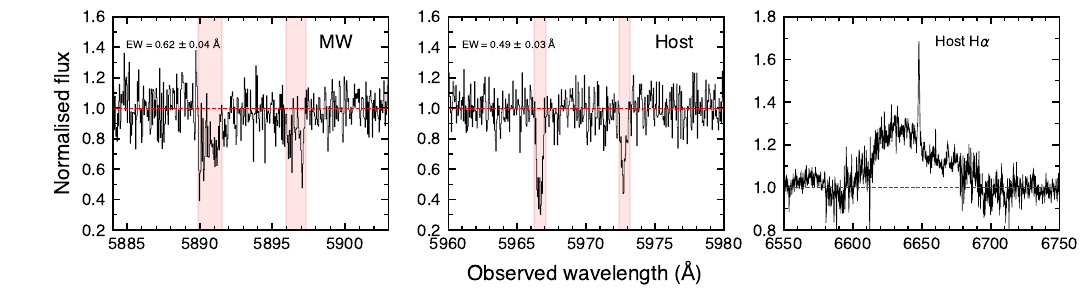}
    \caption{SALT high-resolution spectrum of \SN{}. \textit{Left:} \ion{Na}{i} D absorption doublet due to Milky Way reddening. \textit{Centre:} same as previous, but due to the host galaxy. The shaded regions indicate the range over which EW(D$_1$+D$_2$) was calculated. \textit{Right:} region around H$\alpha$ showing the narrow host line and the broad blue-shifted emission from the SN. } \label{fig:NaID}
\end{figure*}

The MW dust reddening towards the SN line of sight has $E(B-V)=0.062\pm0.002$~mag according to the dust maps of \citet{SF2011}. Using our high-resolution SALT spectrum (Section \ref{subsec:spectra}) we are able to obtain an independent estimate of the reddening due to both the MW and the host galaxy. The left and centre panels of Figure \ref{fig:NaID} show the \ion{Na}{i}~D absorption doublets for the MW and the host galaxy, respectively. We measure the equivalent width (EW) for each of the \ion{Na}{i}~D lines by integrating the absorption profile in a user-specified range. The summed EW for the doublet is converted to a value for $E(B-V)$ using Equation\footnote{$\log_{10}(E(B-V)/\mathrm{mag}) = 1.17\times\mathrm{EW}(\mathrm{D_1}+\mathrm{D_2})/\AA-1.85\pm0.08$} 9 in \citet{Poznanski2012}. For the MW we measure $E(B-V)=0.075\pm0.016$~mag, which agrees with the \citet{SF2011} value within the errors. We note, however, that the MW profile appears to show evidence for two absorption systems, both of which we include in the EW calculation, so our measured value may be an overestimate. For the host galaxy we measure $E(B-V)=0.053\pm0.011$~mag, although we also caution that the relations in \citet{Poznanski2012} were only derived for the MW and may not apply to external galaxies in which the reddening law is likely to be different. Adopting a MW reddening law with $R_V=3.1$ for ESO 501- G 001, the total (MW+host galaxy) visual extinction is therefore $A_V=0.40\pm0.06$~mag. 

The region around H$\alpha$ in our SALT high-resolution spectrum (right panel of Figure \ref{fig:NaID}) shows a narrow host emission line superposed on a broader, blueshifted emission line from the SN. We fit the two lines with a two-component model comprising a Gaussian for the narrow host line and a Lorentzian for the blueshifted component. The host line has a wavelength of $6647.97\pm0.02~\AA$ corresponding to a redshift of $z=0.012979(5)$. The host \ion{Na}{i}~D absorption lines are also at this redshift. These lines are at a systemic velocity of $+85.9\pm1.5$~km~s$^{-1}$ compared to the host redshift, and we interpret this as due to the rotational velocity of the galaxy at the SN location. We discuss the nature of the broad blueshifted component in detail in Section \ref{subsec:flash}.

\begin{table}
\caption{Fundamental parameters of \SN{} and its host galaxy}
\label{tab:SN_gal}
\centering
\begin{tabular}{lc}
\hline\hline
RA (J2000) & $10^\mathrm{h}29^\mathrm{m}36.43^\mathrm{s}$ \\ 
Dec. (J2000) & $-24^\circ06^{\prime}33.61^{\prime\prime}$ \\
Last non-detection (MJD) & 60814.82\\
First detection (MJD) & 60817.05 \\
Explosion epoch (MJD) & $60816.0\pm0.1$\\
Peak time in $r$-band (MJD) & $60827\pm0.5$ \\
Host galaxy & ESO 501- G 001 \\
Morphology & SAB(s)d \\
Host redshift ($z$) & 0.012689(7) \\
SN redshift ($z$) & 0.012979(5) \\
Distance modulus ($\mu$) & $33.54\pm0.09$ \\
Distance (Mpc) & $51.0\pm2.0$ \\
$E(B-V)_\mathrm{MW}$ (mag) & $0.075\pm0.016$ \\
$E(B-V)_\mathrm{host}$ (mag)& $0.053\pm0.011$ \\
$E(B-V)_\mathrm{total}$ (mag) & $0.128\pm0.019$ \\
\hline
\end{tabular}
\end{table}

\section{Photometric properties}\label{sec:photometric_properties}
\subsection{Precursor}\label{subsec:precursor}
\begin{figure*}
    \centering
    \includegraphics[width=\textwidth]{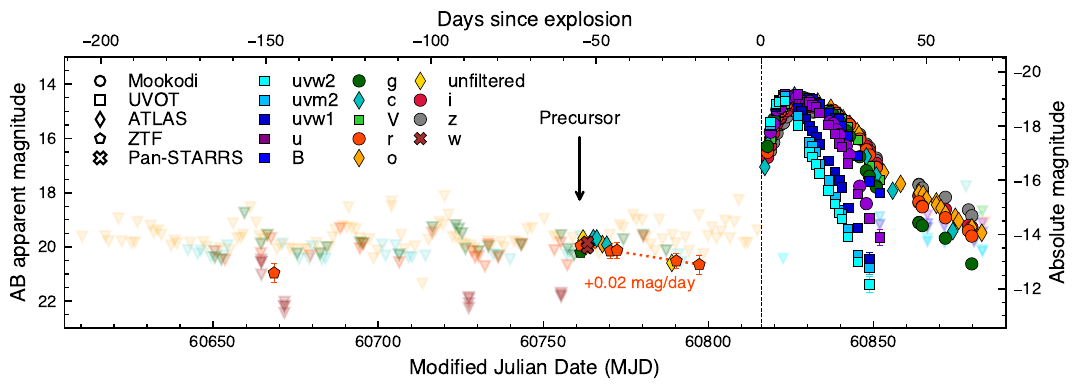}
    \caption{Multi-band photometry of \SN{} from -210 to 75 days post-explosion. The explosion epoch of MJD 60816.0 is indicated by the vertical dashed line, while upper limits are denoted by upside-down triangles.}
    \label{fig:LC_all}
\end{figure*}

\begin{figure}
    \centering
    \includegraphics[width=0.95\columnwidth]{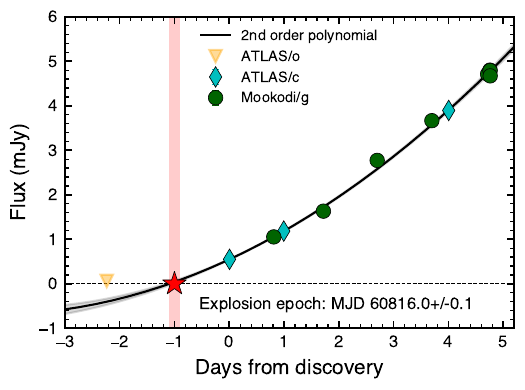}
    \caption{Second-order polynomial fit to our early $c$ and $g$-band photometry to constrain the explosion epoch. The red star indicates the derived explosion epoch, while the vertical red region denotes the 1$\sigma$ uncertainty. Upside-down triangles indicate the most recent pre-explosion ATLAS upper limits. }
    \label{fig:t_explosion}
\end{figure}

Figure \ref{fig:LC_all} presents all photometry of \SN{} spanning 210 days pre-explosion to 75 days post-explosion. To determine the explosion epoch we fitted the early (${<}5$ days post-discovery) rising light curve in the $c$ and $g$ bands with a second order polynomial (Figure \ref{fig:t_explosion}) and converged on an explosion epoch of MJD $60816.0\pm0.1$, which occurred one day prior to the $c$-band discovery. In Figure \ref{fig:LC_all} we do not show archival ATLAS, ZTF, or Pan-STARRS data prior to the range considered here as these data comprise only upper limits. Around 55 days prior to explosion there is a clear precursor event which we regard as robust due to the simultaneous, independent detections by ATLAS, ZTF, and Pan-STARRS. The 5 ZTF $r$-band detections show a monotonic decline, with a decrease in brightness of $\Delta m\approx0.7$~mag between $-55$ and $-19$ days which equates to a decline rate of 0.02~mag~day$^{-1}$. The decline is also clearly seen in the ATLAS unfiltered observations. The earlier, single $r$-band ZTF detection at ${\sim}-150$ days is uncertain due to there being only a single detection.

Assuming blackbody emission, we can use the $g-r$ colour at $-55$ days to estimate the effective temperature and bolometric luminosity of the precursor. Taking into account host and MW extinction (Table \ref{tab:SN_gal}), the observed $g-r$ colour of $0.26\pm0.19$~mag corresponds to a temperature of $8200\pm1800$~K and a blackbody luminosity of ${\approx}10^{41}$~erg~s$^{-1}$. The total radiated energy during the precursor is ${\approx}2\times10^{47}$~erg, assuming a constant temperature evolution. 

Prior to \SN{}, only three SNe Ibn had been detected with precursor emission \citep[SNe 2006jc, 2019uo, and 2023fyq;][]{Pastorello2007,Foley2007,Strotjohann2021,Brennan2024,Dong2024}. In the case of SN 2006jc the precursor was detected as a 9-day-long outburst that occurred two years before explosion with a peak brightness of $M_R{\approx}-14.1$~mag. The precursor to SN~2019uo started ${\sim}340$ days before explosion and was observed over 35 days with an average brightness of $M_r{\approx}-13$~mag. SN~2023fyq showed a markedly different precursor behaviour, with a relatively constant brightness between $-10$ to $-12$~mag between $-1300$ and $-100$ days which was followed by a slow brightening up to ${-15}$~mag before explosion. The \SN{} precursor reached $M_r\approx-13.6$~mag and is more similar to SNe 2006jc and 2019uo in terms of brightness and light curve behaviour as no rising phase similar to that in SN~2023fyq is seen. The timing of the precursor to \SN{} is quite different to those in SNe 2006jc and 2019uo, however, since it begins only two months before explosion rather than 1 to 2 years. We discuss the nature of the precursor further in Section \ref{subsec:precursor}.
  
\subsection{Light curves and colour evolution}
\begin{figure}
    \centering
    \includegraphics[width=\linewidth]{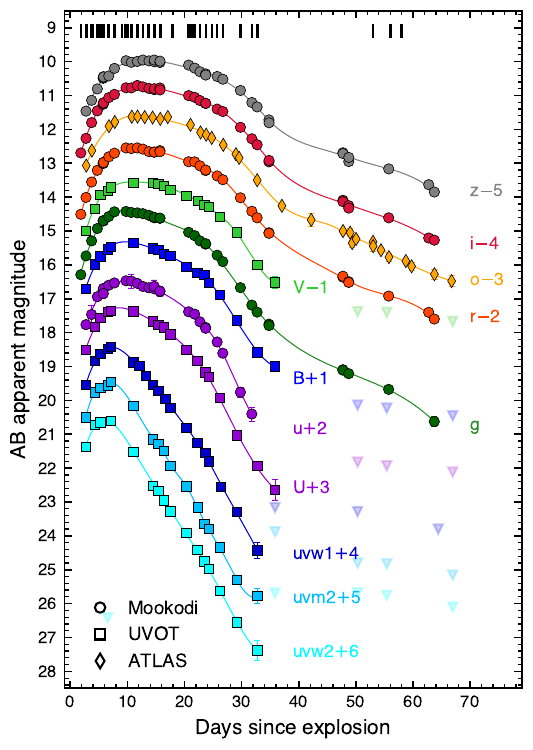}
    \caption{UV/optical light curves of \SN{} along with smooth cubic spline fits to each band. Due to the poor sampling in the ATLAS $c$ band we vertically shift the $V$-band model until a satisfactory fit is obtained. Upper limits are shown as upside-down triangles and are at the $3\sigma$ ($5\sigma$) level for UVOT (ATLAS) data. Vertical markers indicate the times of our spectroscopic observations. }
    \label{fig:LC_shifted}
\end{figure}

We present the UV/optical light curves for \SN{} in Figure \ref{fig:LC_shifted} along with cubic spline fits to each observing band. The parameters derived from our light curve fits include the peak time, peak brightness, the average rise rate until peak ($\Delta m_\mathrm{rise}/\Delta t_\mathrm{rise}$) and the decline within 15 days of peak ($\Delta m_{15}$), as shown in Table \ref{tab:LC_peaks}. All light curves reached a similar peak absolute magnitude of ${\approx}{-19}$ but with the peak time reached earlier in the UV bands (6.1 days in $uvw2$) compared to the redder bands (13.0 days in $z$), reflecting the rapid temperature evolution. Following peak brightness the UV bands declined faster compared to the redder bands, with $\Delta m_{15}=3.54\pm0.03$~mag in $uvw2$ versus $0.69\pm0.22$ in $z$. 

\begin{figure}
    \centering
    \includegraphics[width=0.95\columnwidth]{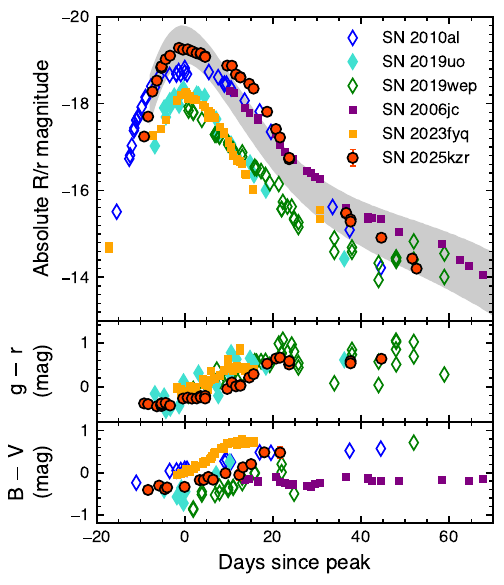}
    \caption{Comparison of $R/r$-band absolute magnitude light curves of a number of SNe Ibn. We include three SNe showing prominent flash features: SNe~2010al \citep{Pastorello2015}, 2019uo \citep{Gangopadhyay2020}, and 2019wep \citep{Gangopadhyay2022}; and three with confirmed precursors: SNe~2006jc \citep{Pastorello2007}, 2019uo \citep{Strotjohann2021}, and 2023fyq \citep{Dong2024}. We show the Ibn template light curve from \citet{Hosseinzadeh2017} in grey. All light curves are corrected for total extinction. The lower panels present the extinction-corrected $g-r$ and $B-V$ colour evolution.}
    \label{fig:absmag_comparison}
\end{figure}

In Figure \ref{fig:absmag_comparison} we compare the $R/r$-band absolute magnitude light curves and the $g-r$ and $B-V$ colour curves for a number of SNe Ibn. The comparison sample includes three objects showing flash-ionised features (SNe~2010al, 2019uo and, 2019wep) and the three SNe Ibn with confirmed precursors (SNe~2006jc, 2019uo, and 2023fyq). We correct each light curve for total extinction (Table \ref{tab:SN_gal}). SN 2019uo is the only object, prior to \SN{}, showing flash features and a precursor. \SN{} reached a peak $r$-band brightness of $M_r={-19.26}\pm0.09$ mag, which agrees closely with the peak brightness of $-19.47^{-0.32}_{+0.54}$ mag for the Type Ibn template light curve derived by \citet{Hosseinzadeh2017}. The overall evolution also appears to match the template quite closely, although \SN{} showed a shallower decline in the first 15 days from peak compared to the template. A steeper decline post-peak is also seen for SNe 2019uo, 2019wep, and 2023fyq. The most similar object in terms of luminosity and light-curve shape is SN~2010al, which also showed prominent flash features \citep{Pastorello2015,Chugai2022}. 

The colour evolution of \SN{} does not stand out within the comparison sample. After a short red-to-blue evolution during the rising phase of the light curve, \SN{} underwent a blue-to-red evolution of ${\approx}1$~mag between $-6$ and 21 days post-peak. Thereafter, the $g-r$ and $B-V$ colours showed an abrupt flattening which coincided with a shallower decay seen in the late-time $r$-band light curve. Assuming that the main light curve peak was powered by CSM interaction, the lack of a continued reddening in the colour evolution may indicate a new energy source taking over after 21 days post-peak.

\begin{table*}
\caption{Light curve parameters derived from cubic spline fits.}
\label{tab:LC_peaks}
\centering
\begin{tabular}{lccccc}
\hline\hline
Band & \makecell{Peak time \\ (days)\tablefootmark{a}} & \makecell{Peak apparent \\magnitude} & \makecell{Peak absolute \\magnitude\tablefootmark{b}} & \makecell{$\Delta m_\mathrm{rise}/\Delta t_\mathrm{rise}$ \\(mag~day$^{-1}$)\tablefootmark{c}} & \makecell{$\Delta m_{15}$ \\(mag)\tablefootmark{d}} \\
\hline
$uvw2$ & $6.1\pm0.1$ & $14.57\pm0.02$ & $-18.97\pm0.02$ & $-0.24\pm0.01$ & $3.54\pm0.05$ \\
$uvm2$ & $7.2\pm0.3$ & $14.48\pm0.02$ & $-19.06\pm0.02$ & $-0.23\pm0.02$ & $3.62\pm0.15$ \\
$uvw1$ & $7.3\pm0.1$ & $14.44\pm0.02$ & $-19.10\pm0.02$ & $-0.24\pm0.01$ & $2.80\pm0.05$ \\
$U$ & $8.6\pm0.2$ & $14.27\pm0.02$ & $-19.27\pm0.02$ & $-0.21\pm0.01$ & $1.88\pm0.07$ \\
$u$ & $9.8\pm1.9$ & $14.47\pm0.08$ & $-19.07\pm0.08$ & $-0.19\pm0.05$ & $1.71\pm0.34$ \\
$B$ & $9.8\pm0.5$ & $14.34\pm0.03$ & $-19.20\pm0.03$ & $-0.19\pm0.02$ & $1.27\pm0.12$ \\
$g$ & $9.8\pm0.3$ & $14.41\pm0.02$ & $-19.13\pm0.02$ & $-0.24\pm0.01$ & $1.15\pm0.07$ \\
$V$ & $12.1\pm1.7$ & $14.57\pm0.05$ & $-18.97\pm0.05$ & $-0.15\pm0.03$ & $1.16\pm0.25$ \\
$r$ & $10.8\pm0.5$ & $14.56\pm0.02$ & $-18.98\pm0.02$ & $-0.22\pm0.01$ & $0.83\pm0.06$ \\
$o$ & $12.0\pm0.4$ & $14.59\pm0.03$ & $-18.95\pm0.03$ & $-0.16\pm0.01$ & $0.89\pm0.04$ \\
$i$ & $11.8\pm0.5$ & $14.74\pm0.02$ & $-18.80\pm0.02$ & $-0.20\pm0.01$ & $0.78\pm0.07$ \\
$z$ & $13.0\pm2.6$ & $14.97\pm0.04$ & $-18.57\pm0.04$ & $-0.15\pm0.04$ & $0.70\pm0.25$ \\
\hline
\end{tabular}
\tablefoot{\\
\tablefoottext{a}{Days since the explosion epoch of MJD $60816.0\pm0.1$.} \\
\tablefoottext{b}{Calculated assuming a distance modulus of $\mu=33.54\pm0.09$. We do not include the distance modulus uncertainty in our error estimates, nor do we correct for extinction.} \\
\tablefoottext{c}{Rise rate from first detection until peak.}\\
\tablefoottext{d}{Decline within 15 days of peak.}
}
\end{table*}

\subsection{Bolometric light curve}\label{subsec:bol}
\begin{figure*}
\centering
\begin{subfigure}[b]{0.48\textwidth}
    \includegraphics[]{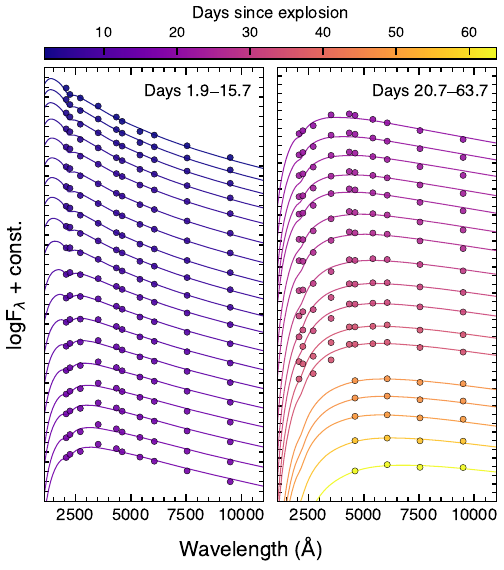}
    \caption*{} 
\end{subfigure}
\hfill
\begin{subfigure}[b]{0.48\textwidth}
    \includegraphics[]{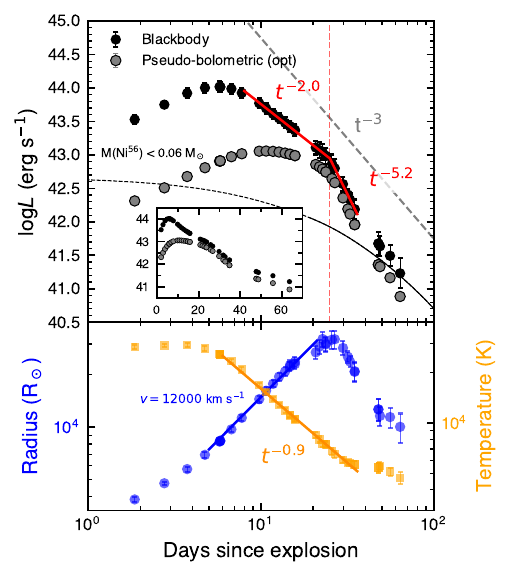}
    \caption*{} 
\end{subfigure}
\vspace{-1em}
\caption{
Left: Interpolated SEDs from our light curve fits ranging from 1.9 to 63.7 days post-explosion, along with the best-fitting, reddened blackbody SEDs. We shift each SED vertically for clarity. The last five epochs comprise $griz$ photometry only. Right: Bolometric light curves (top) along with the temperature and radius evolution (bottom) as derived from our blackbody fits. The inset presents the bolometric light curves on a linear scale. We show a broken power-law fit to the blackbody luminosity light curve in red. The vertical dashed line denotes the break time of 25~days from the fit. The grey line highlights the $t^{-3}$ decline expected during the steep decay phase in the SN Ibn light curve model of \cite{Maeda2022}. The dashed and solid black lines indicate the contribution from the radioactive decay of 0.06~$M_\odot$ of \textsuperscript{56}Ni assuming $M_\mathrm{ej}=2~M_\odot$ and $E_\mathrm{K}=10^{51}$~erg. Diffusion becomes important when the dashed line transitions to the solid line. The blue solid line indicates a photosphere expanding at 12000~km~$^{-1}$ between 5 and 25 days. We also show a power-law fit to the temperature evolution between 5 and 40 days. }
\label{fig:Lbol}
\end{figure*}

To construct a bolometric light curve we extracted UV/optical spectral energy distributions (SEDs) by interpolating our cubic spline fits (Figure \ref{fig:LC_shifted}) to the times of our $g$-band detections. \SN{} was not detected in the UV bands after 40 days post-explosion, so at these later times we extracted $griz$ SEDs. At each epoch we fit a reddened blackbody function by comparing the observed SED with synthetic photometry calculated using the UVOT and Mookodi filter transmission curves and the \verb|pysynphot| software package. We excluded the ATLAS photometry due to their broad filter bandpasses. We also excluded the Mookodi $u$-band data\footnote{At times ${<20}$ days we noticed that our Mookodi $u$-band magnitudes were systematically below the blackbody predictions from the \Swift{}/UVOT and $griz$ photometry. After thorough investigation we concluded that the discrepancy arose due to the extremely hot SN spectrum. We attempted to perform a colour correction to account for the extreme spectrum at these times, but this was unsuccessful due to the small number of $u$-band calibration stars in the field-of-view.} from our fits. Figure \ref{fig:Lbol} demonstrates that the observed SEDs are well-described by a blackbody function, especially during the first 20 days when the blackbody temperature was high (${>}10000$~K). At later times the fit is still good although there is a larger discrepancy in the UV bands. 

We derive two bolometric light curves: one for which we assume a perfect blackbody with the luminosity calculated using the Stefan Boltzmann Law; and a "pseudo-bolometric" light curve derived by integrating the observed SED across the optical wavelength range which encompasses the $griz$ bands. The discrepancy between the two light curves (Figure \ref{fig:Lbol}) at times $t<10$~days is explained by a large fraction of the emission originating in the UV regime due to the high temperatures. The blackbody luminosity peaks at ${\sim}10^{44}$~erg~s$^{-1}$ at 5.8 days post-explosion, which is ${\sim}2$~days after the peak temperature of ${\sim}29000$~K is reached. By contrast, the peak pseudo-bolometric luminosity is an order of magnitude fainter than the blackbody luminosity and peaks later at ${\sim}10$~days, which is around the time of the $g$- and $r$-band peak times (Table \ref{tab:LC_peaks}). The blackbody radius increases from an initial radius of ${\sim}4000~R_\odot$ at 1.9 days to a peak of ${\sim}32000~R_\odot$ at 22.7 days, before declining. The expansion rate of the blackbody photosphere before peak is ${\approx}12000$~km~s$^{-1}$, although during the first four days we observe a slower expansion speed closer to ${\approx}8000$~km~s$^{-1}$. The peak in radius coincides with a steepening decline in the bolometric light curve. \citet{Maeda2022} investigated the bolometric light curves of SNe Ibn and found that they generally show a shallow decay with $L\propto t^{-1}$ that transitions to a steeper decay with $L\propto t^{-3}$. A piecewise broken power-law fit to the blackbody light curve demonstrates that the initial decline post-peak follows $L\propto t^{-2.0}$ which steepens to a decline with $L\propto t^{-5.2}$ at 25 days, with the steepening coinciding with the peak in the radius evolution. These observed decay rates are steeper than the empirical decay rates, although we note that the bolometric light curves in \citet{Maeda2022} may not have included the UV emission, which clearly results in a steeper early decay for \SN{}. 

In the light curve model developed by \citet{Maeda2022}, the steepening of the light curves is explained by a transition of the forward shock from a cooling to an adiabatic regime, without there being a change in the density distribution of the CSM. The results of their modelling show that SNe Ibn have steeper density profiles compared to normal stellar winds, with $\rho_\mathrm{CSM}\propto r^{-3}$, indicating a rapid increase in the mass-loss rate prior to explosion. The decay rate of $t^{-4.5}$ we measure for \SN{} might indicate an even steeper density distribution, although this is difficult to model since there are no self-similar solutions to the ejecta-CSM interaction problem when $s>3$ \citep{Chevalier1982}. 

Interestingly, the temperature decline follows a very tight power-law with an index of $\alpha=0.89\pm0.01$ ($T\propto t^{-\alpha}$) that does not deviate even after the light curve steepens at 25 days. Further investigation is needed to determine the physical significance and origin of such a tight power-law. The last few epochs of photometry after 40 days show that the temperature did not continue with its power-law decline, but instead plateaued at around 5500~K before declining again, consistent with the flattening seen in the colour curves (Figure \ref{fig:absmag_comparison}). One possibility is that radioactivity powers the bolometric light curve at these times. Similar to \citet{Maeda2022}, we use our light curve to place constraints on the mass of \textsuperscript{56}Ni synthesised in the explosion. Using a canonical explosion energy of $E_\mathrm{K}=10^{51}$~erg and an ejecta mass of $M_\mathrm{ej}=2~M_\odot$, 0.06~$M_\odot$ of \textsuperscript{56}Ni can account for the late-time behaviour (Figure \ref{fig:Lbol}). With larger ejecta masses, lower \textsuperscript{56}Ni masses would be required, so 0.06~$M_\odot$ is likely an upper limit. \citet{Maeda2022} computed only upper limits for all of the objects in their sample, and found that the \textsuperscript{56}Ni masses were in general lower than those of SESNe, and therefore point towards a different progenitor. A value of 0.06~$M_\odot$, however, is certainly on the lower-end of the distribution for SESNe (see their Figure 8).  We discuss this further in Section \ref{subsec:progenitor}.

\subsection{Light curve modelling}\label{subsec:modelling}
The light curves of SNe Ibn are generally thought to be powered by the ejecta-CSM interaction, and have been modeled with a combination of CSM-only and CSM+radioactive decay models \citep{Gangopadhyay2020,BenAmi2023,Pursiainen2023,Dong2024,Gangopadhyay2025}. Here we use a combination of the CSM-interaction model of \citet{Jiang2020} and the \textsuperscript{56}Ni decay formalism of \citet{Chatzopoulos2012}, as used by \citet{Pellegrino2022} and \citet{Dong2024}. 

The \citet{Jiang2020} model generalises the self-similar solution to the interaction of a uniformly expanding SN ejecta ($\rho_\mathrm{ej}\propto r^{-\delta,-s}$) and a stationary CSM (with $\rho\propto r^{-s}$) for $0\leq s\leq 2$, and follows on from the original work of \citet{Chevalier1982}. The free parameters in the combined model are the ejecta velocity ($v_\mathrm{ej}$), ejecta mass ($M_\mathrm{ej}$), CSM mass ($M_\mathrm{CSM}$), inner radius of the CSM ($R_0$), density at the inner radius ($\rho_0$), radiation efficiency ($\epsilon$), density profile of the CSM ($s$), density profiles of the inner ($\delta$) and outer ejecta ($n$), the mass of \textsuperscript{56}Ni synthesized in the explosion ($M_\mathrm{Ni}$), and the gamma-ray opacity ($\kappa_{\gamma}$). To reduce the number of free parameters, we adopt the same assumptions as in \citet{Pellegrino2022}: the inner ejecta have a shallow density profile with $\delta=1$ and a steep outer density profile with $n=10$; the CSM is shell-like with $s=0$; and the optical opacity is fixed at $\kappa_\mathrm{opt}=0.1$~cm$^2$~g$^{-1}$, as appropriate for a He-rich composition. We choose to fix the ejecta velocity $v_\mathrm{ej}$ at 5000~km~s$^{-1}$, consistent with the velocity of the ejecta features in our spectra (Section \ref{subsec:ejecta}). We performed global parameter optimization using the differential evolution algorithm in \texttt{SciPy}, followed by refinement with the Nelder--Mead simplex method to minimize the chi-squared of the fit. 

\begin{figure}
    \centering
    \includegraphics[width=\linewidth]{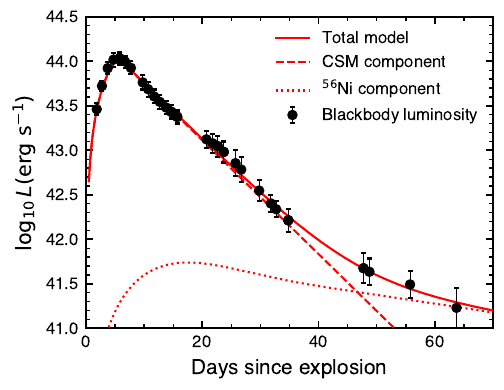}
    \caption{Blackbody luminosity along with the best-fit ejecta-CSM interaction model. We show separately the CSM (dashed) and \textsuperscript{56}Ni (dotted) components.}
    \label{fig:bol_model}
\end{figure}

The best-fit model (Figure \ref{fig:bol_model} and Table \ref{tab:bol_fit}) shows good agreement with the bolometric light curve, with the radioactivity-powered component accounting for the shallower late-time decay. The model does not, however, account for the shallow-to-steep decay with a break time of 25~days more clearly visible in Figure \ref{fig:Lbol}. A Markov chain Monte-Carlo (MCMC) exploration shows that a wide range of parameters can accommodate the data (Table \ref{tab:bol_fit}), making it difficult to draw strong conclusions on the physical parameters. Although the shell-like density profile (with $s=0$) used in this model is markedly different from the much steeper density profile (with $s=3$) suggested by \citet{Maeda2022} for SNe Ibn, we note that this model finds a large amount of CSM confined within only ${\sim}80$~au (${\sim}1.2\times10^{15}$~cm), which implies a steep drop off in the density beyond this radius. Although there is a lack of consensus on physical models of SN Ibn light curves, if CSM interaction is powering the \SN{} light curves our results suggest a CSM density that drops steeply.    

\begin{table*}
\centering
\caption{Best-fit and MCMC-derived model parameters.}
\label{tab:bol_fit}
\begin{tabular}{lccccccc}
\hline
 & $M_\mathrm{ej}$ ($M_\odot$) &
$M_\mathrm{CSM}$ ($M_\odot$) &
$R_0$ (au) &
$\log_{10}(\rho_0/\mathrm{g\,cm^{-3}})$ &
$\epsilon$ &
$M_\mathrm{Ni}$ ($M_\odot$) &
$\kappa_\gamma$ ($\mathrm{cm^2\,g^{-1}}$) \\
\hline
Best-fit &
1.5 &
1.6 &
82 &
-10.0 &
0.52 &
0.02 &
0.05 \\
MCMC &
$1.6^{+3.5}_{-1.1}$ &
$1.7^{+1.0}_{-0.9}$ &
$86^{+46}_{-44}$ &
$-9.8^{+0.4}_{-0.4}$ &
$0.52^{+0.30}_{-0.24}$ &
$0.02^{+0.02}_{-0.02}$ &
$0.20^{+2.53}_{-0.19}$ \\
\hline
\end{tabular}
\end{table*}

\section{Spectroscopic properties}\label{sec:spectroscopic_properties}
\subsection{Overall evolution}
\begin{figure}
    \centering
    \includegraphics[width=\linewidth]{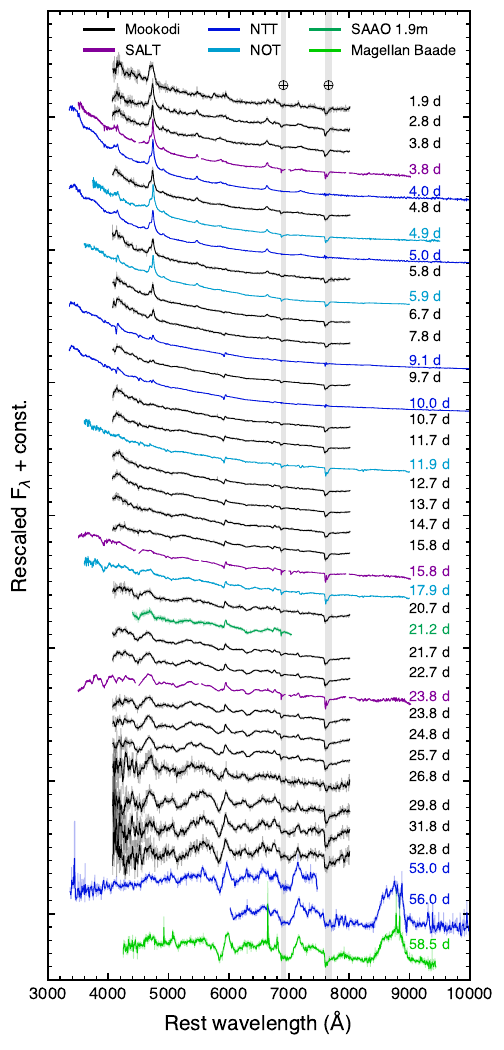}
    \caption{Low-resolution spectra of \SN{} spanning 1.85 d to 58.49 d post-explosion. We smooth the Mookodi, SAAO 1.9~m and the final three spectra using a Savitzky-Golay filter. Spectra are not corrected for extinction.}
    \label{fig:all_spec}
\end{figure}

Figure \ref{fig:all_spec} presents all of our low-resolution ($R{<}2000$) spectra for \SN{} spanning 1.9 to 58.5 days post-explosion. The spectral sequence shows in exquisite detail the disappearance and emergence of key spectral features. 
During the first ten days of evolution, \SN{} showed prominent emission lines of ionised species including \ion{He}{ii}, \ion{N}{iii}, \ion{C}{iii}, \ion{C}{iv}, and \ion{N}{iv} superposed over a hot, blue continuum. Such emission lines have been seen in other flash-ionised SNe \citep[e.g. SNe 1998S, 2013cu, and 2023ixf;][]{Fassia2001,Galyam2014,Bostroem2023} and are believed to originate in unshocked CSM that has been ionised by radiation from the ejecta-CSM interaction shock. We refer to this phase as the `flash phase' henceforth. A total of seven SNe Ibn have been detected with flash features \citep[SNe 2010al, 2019uo, 2019wep, 2019cj, 2023emq, 2023xgo, 2024acyl;][]{Pastorello2015,Gangopadhyay2020,Gangopadhyay2022,Wang2024,Pursiainen2023,Gangopadhyay2025,Cai2026,Yamanaka2025}, although SNe 2023emq and 2023xgo are regarded as transitional Ibn/Icn objects. \SN{} is therefore the eighth SN Ibn to show flash-ionised features. Our dataset comprising 15 spectra obtained during the flash phase makes this by far the best-observed flash-ionised SN Ibn to date. Before the flash lines fully disappear and while the continuum is still hot ($T{\sim}23000$~K at 6.7 days) we see the emergence of intermediate-velocity \ion{He}{i} P Cygni lines that are characteristic of the SN Ibn class. We refer to this phase as the `Helium P Cygni phase'. In the spectrum at 15.8 days we start to see the emergence of broader features typical of SN spectra during the photospheric phase and which probe the SN ejecta. These broad features are most distinct in our SALT spectrum at 23.8 days, so we refer to this phase as the `ejecta phase'. The final phase in the spectral evolution after 50 days shows a strong \ion{Ca}{ii} emission feature at 8600~\AA~that is often seen in the nebular spectra of SESNe. We refer to this phase as the `late-time phase' henceforth. We investigate each of these phases in further detail below.  

\subsection{The flash phase}\label{subsec:flash}

\subsubsection{Early evolution}
\begin{figure*}
    \centering
    \includegraphics[]{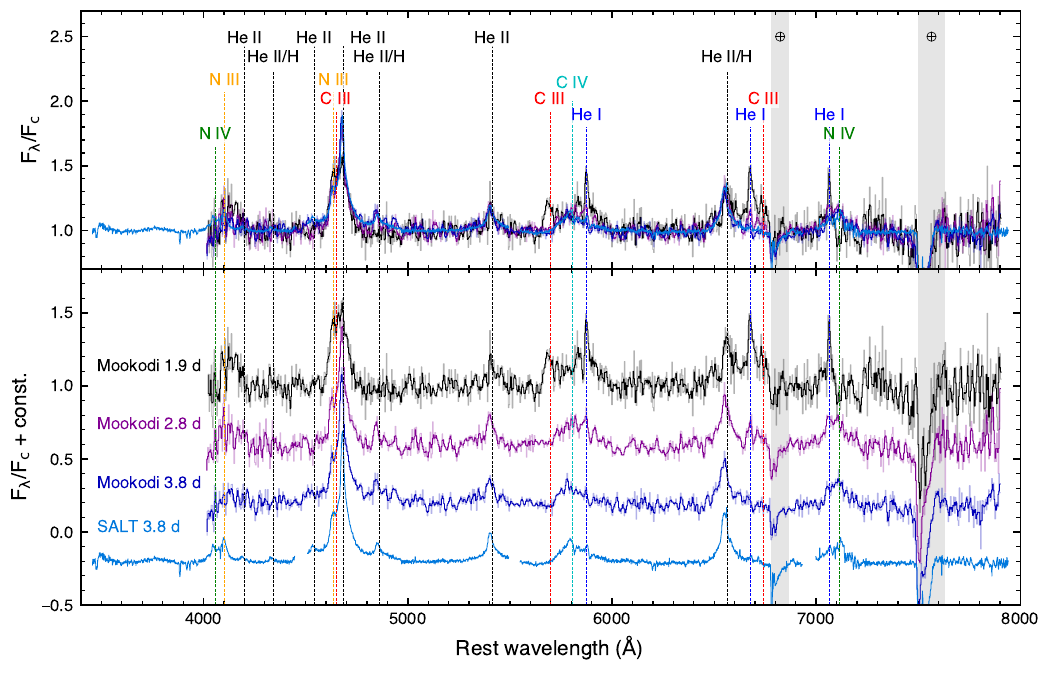}
    \caption{First four continuum-normalised spectra for \SN{} highlighting the most prominent spectral lines. Vertical grey regions denote regions of telluric absorption. }
    \label{fig:first4days}
\end{figure*}

The most dramatic spectral changes occurred within the first four days, as demonstrated in Figure \ref{fig:first4days}. Our first Mookodi spectrum shows strong \ion{He}{i}~$\lambda\lambda5876$, $6678$, and $\lambda7065$ emission lines which rapidly weaken over the first three epochs. Additionally, the \ion{C}{iii}~$\lambda5696$ line completely disappears between the first and second epochs, while the \ion{C}{iii}/\ion{N}{iii} $\lambda4650$ blend also weakens considerably over this period. The \ion{C}{iii}~$\lambda5696$ line is rarely seen in Type Ibn SNe \citep[except for SNe~2023emq and 2023xgo, which are `transitional' Ibn/Icn SNe;][]{Pursiainen2023,Gangopadhyay2025} but is a far more prominent line in SNe Icn which show higher-ionisation C lines in their spectra with weak or absent H and He. Our highest-resolution spectrum during the flash phase is our SALT spectrum at 3.8 days, which coincided with the peak in the photospheric temperature. This spectrum contains a wealth of information which we focus on below.

\subsubsection{A H-free, WN-like spectrum}\label{subsubsec:pickering}
The emission lines in our continuum-normalised SALT spectrum show a striking resemblance to spectra of Wolf-Rayet (WR) stars of the nitrogen subclass (WN), which are known to show strong emission lines of He and N \citep[see][for a review]{Crowther2007}. 
The classification of WN stars is based on the line strengths and ratios of \ion{He}{i-ii} and \ion{N}{iii-V} lines and is meant to reflect an ionisation sequence, with earlier subtypes showing stronger \ion{He}{ii} versus \ion{He}{i} and stronger \ion{N}{iv/v} versus \ion{N}{iii}. 

\begin{figure}
    \centering
    \includegraphics[]{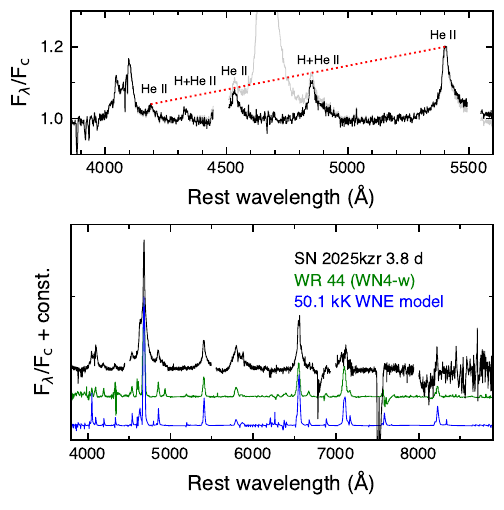}
    \caption{Top panel: SALT spectrum demonstrating the Pickering-decrement method applied to \SN{}. The \ion{He}{ii} $\lambda4686$ complex has been subtracted from the original spectrum (grey). The red dotted line joins the \ion{He}{ii} $\lambda\lambda5411,4200$ lines and reveals a H-free spectrum. Bottom panel: SALT 3.8 day spectrum compared to the spectrum of WR 44 from \citet{Hamann1995} and a 50.1 kK WNE model with $\log (R_\mathrm{t}/R_\odot)=1.1$ \citep{Hamann2004,Todt2015}. The comparison spectra have been normalised to match the height of the \ion{He}{ii}~$\lambda4686$ line and shifted vertically for clarity.}
    \label{fig:WR_comparison}
\end{figure}

A natural question in the case of \SN{} concerns the presence of hydrogen. Our SALT spectrum is noteworthy for the clearly identified lines of the Pickering series, which consists of $n{\rightarrow}4$ transitions of the \ion{He}{ii} ion. Since \ion{He}{ii} is a hydrogenic ion, the even-$n$ transitions happen to coincide with the Balmer lines of hydrogen, albeit with small shifts in wavelength (e.g. 6560 vs 6563~$\AA$ for H$\alpha$) due to the slightly heavier reduced mass for the \ion{He}{ii} ion. The hydrogen abundance in WN stars has been studied extensively via the Pickering decrement method \citep{Conti1983,Smith1996}, which involves comparing the line fluxes of blended H+\ion{He}{ii} lines (even-$n$ Pickering transitions) versus isolated \ion{He}{ii} lines (odd-$n$ transitions). A simple, qualitative classification method devised by \citet[][see their Figure 6]{Smith1996} is to draw a line connecting the peaks of the $\lambda4200$ ($11{\rightarrow}4$), $\lambda4541$ ($9{\rightarrow}4$), and $\lambda5411$ ($7{\rightarrow}4$) lines and determine if the blended $\lambda4340$ ($10{\rightarrow}4$) and $\lambda4861$ ($8{\rightarrow}4$) lines exceed the height of this line. Figure \ref{fig:WR_comparison} demonstrates this method applied to our SALT spectrum. It is clear that the blended H+\ion{He}{ii} lines do not exceed the height of the isolated \ion{He}{ii} lines. This method can be made quantitative by determining the ratio in EW of a H+\ion{He}{ii} line with the geometric mean of pure \ion{He}{ii} lines on either side, minus one, which yields the relative abundance of H to He \citep{Smith1996}. Since the H+\ion{He}{ii} lines are clearly below the line drawn between neighbouring \ion{He}{ii} lines, we compute a hydrogen abundance of zero. This strongly suggests that the dense, flash-ionised CSM of \SN{} is completely hydrogen-free, in line with the commonly-held view that the progenitors of SNe Ibn exploded in a He-rich CSM with little to no hydrogen.   

Figure \ref{fig:WR_comparison} demonstrates that the WN4-w spectrum of WR 44 \citep{Hamann1995} shows an overall similarity with our SALT spectrum. To place this comparison in context, we also examined a large grid of hydrogen-free WNE (WN Early-type) model spectra \citep{Hamann2004,Todt2015} parameterised by the stellar temperature $T_*$ and the transformed radius $R_\mathrm{t}$, both of which determine the relative strengths and EWs of the various emission lines. All models assume a terminal wind velocity of 1600~km~s$^{-1}$. To identify models consistent with our SALT spectrum we devised a simple peak-matching routine that compares the heights of the \ion{He}{ii}~$\lambda\lambda\lambda4686$, $5411$, $6560$, \ion{C}{iv}~$\lambda5801$, and \ion{N}{iv}~$\lambda7115$ lines with those in our SALT spectrum. We found that using relative heights rather than the absolute peak heights yielded better agreement in terms of the observed line profiles, although the model EWs remained larger than those observed. Figure \ref{fig:WR_comparison} presents the best-matching model, which itself resembles the WR 44 spectrum, and serves to illustrate that our spectrum is generally consistent with H-free WN spectra. We note, however, that the steady winds of WR stars differ fundamentally from the dense, likely slower-moving CSM around \SN{} which is ionised through the ejecta–CSM interaction shock.

\subsubsection{Line profiles}\label{subsubsec:line_profiles}
\begin{figure*}
    \centering
    \includegraphics[]{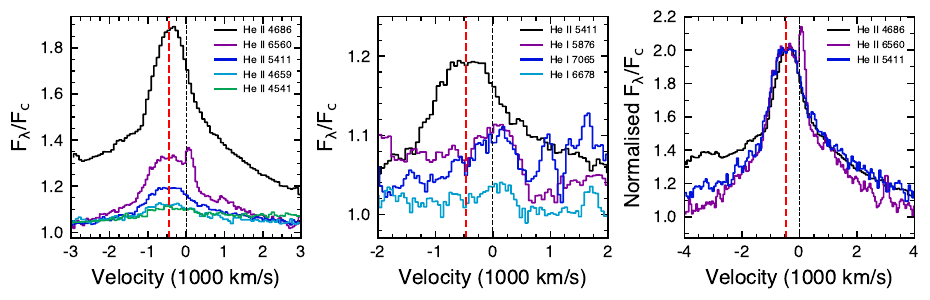}
    \caption{Line profiles for selected lines in our SALT 3.8 day spectrum. The black and red vertical dashed lines indicate velocities of zero and 460~km~s$^{-1}$, respectively. Left: line profiles for \ion{He}{ii} lines where we have assumed \ion{He}{ii} wavelengths for the lines near H$\alpha$ and H$\beta$. Centre: line profiles for \ion{He}{i} lines along with \ion{He}{ii}~$\lambda5411$. Right: line profiles for the three strongest \ion{He}{ii} lines where each line has been normalised to the same height.} 
    \label{fig:SALT_line_profiles}
\end{figure*}

A second striking aspect of our SALT spectrum concerns the line profiles and velocities. Recall that our high-resolution SALT spectrum showed a narrow host H$\alpha$ line at a systemic velocity of $+85.9$~km~s$^{-1}$ compared to the average redshift for the host galaxy (Figure \ref{fig:NaID}). Adopting the redshift of this narrow line as zero velocity henceforth, we see that the \ion{He}{ii} lines are all clearly blueshifted by ${\sim}460$~km~s$^{-1}$ compared to zero velocity (left panel in Figure \ref{fig:SALT_line_profiles}). Besides the \ion{He}{ii}~$\lambda4686$ line, the lines of the Pickering series appear to show flat-topped or even weakly double-peaked line profiles, which is either due to a shell-like geometry or optical depth effects. 

Perhaps an even more striking finding from our line profiles is found when we compare the \ion{He}{ii}~$\lambda5411$ and \ion{He}{i} lines (Figure \ref{fig:SALT_line_profiles}). Although the \ion{He}{i} lines are weak at this epoch, they are clearly identified. However, instead of the line centres being blueshifted in a similar fashion to the higher-ionisation emission lines (e.g. \ion{He}{ii}~$\lambda5411$), they are at zero velocity. 
This points to the \ion{He}{i} lines forming in a different emission region compared to the higher-ionisation lines. The strong \ion{He}{i} lines seen in our very first Mookodi spectrum are also at zero velocity. We return to this in the discussion (Section \ref{subsec:blueshift}).
 
\begin{figure}
    \centering
    \includegraphics[]{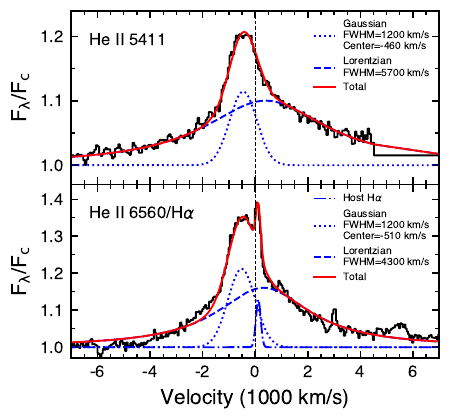}
    \caption{Fits to the \ion{He}{ii}~$\lambda5411$ and \ion{He}{ii}~$\lambda6560$~lines using a model comprising the sum of a broad Lorentzian for the wings and Gaussian for the core. We include an additional narrow Gaussian to account for the narrow host H$\alpha$ line in the \ion{He}{ii}~$\lambda6560$ profile. }
    \label{fig:SALT_line_fits}
\end{figure}

The line profiles of the higher-ionisation emission lines show evidence for broad electron-scattering wings often seen in flash-ionised or interacting SNe \citep{Chugai2001,Smith2008,Galyam2014,Yaron2017}. The normalised line profiles of the three strongest \ion{He}{ii} lines (right panel in Figure \ref{fig:SALT_line_profiles}) demonstrate that the shape of each line is quite similar, although there is a small deficit in flux on the blue side of the \ion{He}{ii}~$\lambda4686$ line near its core which is likely due to the greater optical depth in this line compared to the Pickering series. The similar shape of the broad wings also favours there being little to no hydrogen. If significant hydrogen were present we would expect the wings of the H$\alpha$ line to be significantly weaker and narrower than the \ion{He}{ii} lines due to forming in outer regions where the electron-scattering optical depth is lower. In Figure \ref{fig:SALT_line_fits} we show fits to the \ion{He}{ii}~$\lambda5411$ and \ion{He}{ii}~$\lambda6560$ lines (we assume zero H) where the model we fit comprises the sum of a Gaussian to account for the narrower core of the line and a Lorentzian to account for the broader wings. We choose these two lines as they are isolated and uncontaminated by nearby spectral features. For the \ion{He}{ii}~$\lambda6560$ line we include an additional narrow Gaussian component to account for the host H$\alpha$ contribution. We measure identical FWHM values of 1200~km~s$^{-1}$ for the core component of both lines, with the line centres blueshifted by ${\sim}500$~km~s$^{-1}$. The broad Lorentzian components have FWHM values ranging from 5700 to 4300~km~s$^{-1}$ for the \ion{He}{ii}~$\lambda5411$ and $\lambda6560$ lines, respectively. The centre of each broad component is not blueshifted and is closer to zero velocity which leads to asymmetric line profiles with stronger red wings. Studies of electron scattering wings in interacting SNe and WR stars have shown that an asymmetry to the red is explained by a systematic redshift of the scattering photons in an outflow or wind \citep{Huang2018,Hillier1991}, which is clearly the case here based on the strong blueshift of the flash lines.


\subsubsection{Flash line evolution}\label{subsubsec:EW}
\begin{figure*}
    \centering
    \includegraphics[]{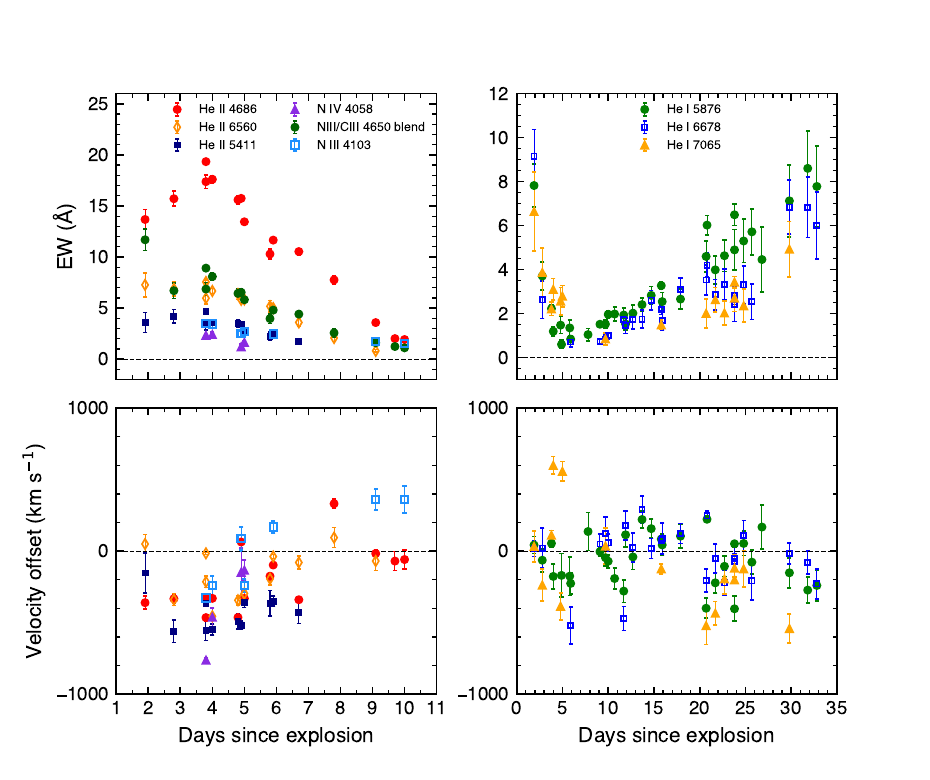}
    \caption{Evolution of the EW (top) and centroid velocity (bottom) for the most prominent flash-ionisation and \ion{He}{i} lines. During the \ion{He}{i} P Cygni phase we only use the emission component when calculating the EW and velocity offset. We only include data points where the EW measurement was three times the RMS noise ($3\sigma$). }
    \label{fig:EW_evolution}
\end{figure*}

In Figure \ref{fig:EW_evolution} we present the evolution of the EW and centroid velocity for the most prominent flash-ionised emission lines as well as the three most prominent \ion{He}{i} lines. To calculate the EW we devised a method which involved integrating the continuum-normalised flux within a $30~\AA$ window centred on the centroid of each line, with uncertainties estimated using the RMS value of the continuum flux. We employed a fixed $30~\AA$ window to avoid including flux from the broad electron-scattered wings. A drawback of this method is that certain lines may be contaminated by nearby spectral features. Nevertheless, it allows us to calculate the EW for multiple lines in an automated, reproducible way that is useful for visualising overall trends. The scatter between measurements of the same line is driven in large part by the uncertainty in the continuum flux level as well as the varying spectral resolution and SNR of our low resolution spectra. 

The strongest line in our flash spectra is \ion{He}{ii}~$\lambda4686$ which rapidly grows in strength to reach a peak at ${\sim}3.8$ days, which is also the time of the peak temperature in the continuum (Figure \ref{fig:Lbol}). The other higher-ionisation lines show similar behaviour in peaking at ${\sim}3.8$ days and then weakening, except for the \ion{N}{iii}/\ion{C}{iii} blend just blueward of \ion{He}{ii}~$\lambda4686$. This feature is strong in our first spectrum and weakens quickly thereafter. We note that this feature was also strong in the first spectrum of SN~2010al (see Figure \ref{fig:spectra_comparison} below). Similarly, the \ion{C}{iii}~$\lambda5696$ line (not shown) disappears rapidly after being visible only in our first spectrum. For the \ion{He}{i} lines we calculate the EW using only the emission peak (i.e. we exclude the blueshifted absorption component), and include all spectra up until 35 days post-explosion. The \ion{He}{i} lines were strongest in our very first spectrum at 1.9 days, and weakened rapidly within the first five days during which the photosphere reached its peak temperature. Thereafter, the lines continued to grow gradually in strength. 

The timing of the disappearance of each line is dependent on the peak strength, with the \ion{N}{iv}~$\lambda4058$ line disappearing by ${\sim}6$~days, followed by \ion{He}{ii}~$\lambda5411$ at ${\sim}8$ days, \ion{He}{ii}~$\lambda6560$ at ${\sim}9$ days and finally the \ion{N}{iii}/\ion{C}{iii} blend along with \ion{He}{ii}~$\lambda4686$ and \ion{N}{iii}~$\lambda4103$ between 10 and 12 days. The fact that most of the higher-ionisation lines reach peak strength around the time of peak photospheric temperature indicates that the ionisation level of the CSM was increasing up until this time. The flash-ionised Type II SN~2023ixf showed similar behaviour, and \citet{Smith2023} used this to argue that the photoionisation of the CSM was powered by emission from the ongoing CSM interaction shock rather than a sudden flash of ionisation due to shock breakout from the stellar surface. The same scenario likely applies here. Furthermore, although flash lines can weaken and disappear due to recombination and cooling, we do not see a corresponding increase in strength of the \ion{C}{iii}, \ion{N}{iii}, and \ion{He}{i} lines as the higher-ionisation lines fade, which suggests that the disappearance of the lines was caused by the forward shock sweeping through the unshocked CSM \citep{Smith2023}.  

The velocity evolution of the line centres is particularly interesting, as shown in the lower panels of Figure \ref{fig:EW_evolution}. The flash lines appear to show a general blue-to-zero velocity evolution, while the \ion{He}{i} lines are scattered about zero velocity throughout their evolution. We note here that our SALT spectra had the highest spectral resolution of all our spectra, with $\Delta\lambda=5.7$~\AA~($R{\sim}880$) at 4500~\AA~and a corresponding velocity resolution of $\Delta v=340$~km~s$^{-1}$. This reduced ability to distinguish small shifts in velocity may explain the large scatter between different lines and epochs. We therefore caution against overinterpretation of these results. 

\subsubsection{Spectral comparison}
\begin{figure*}
    \centering
    \includegraphics[]{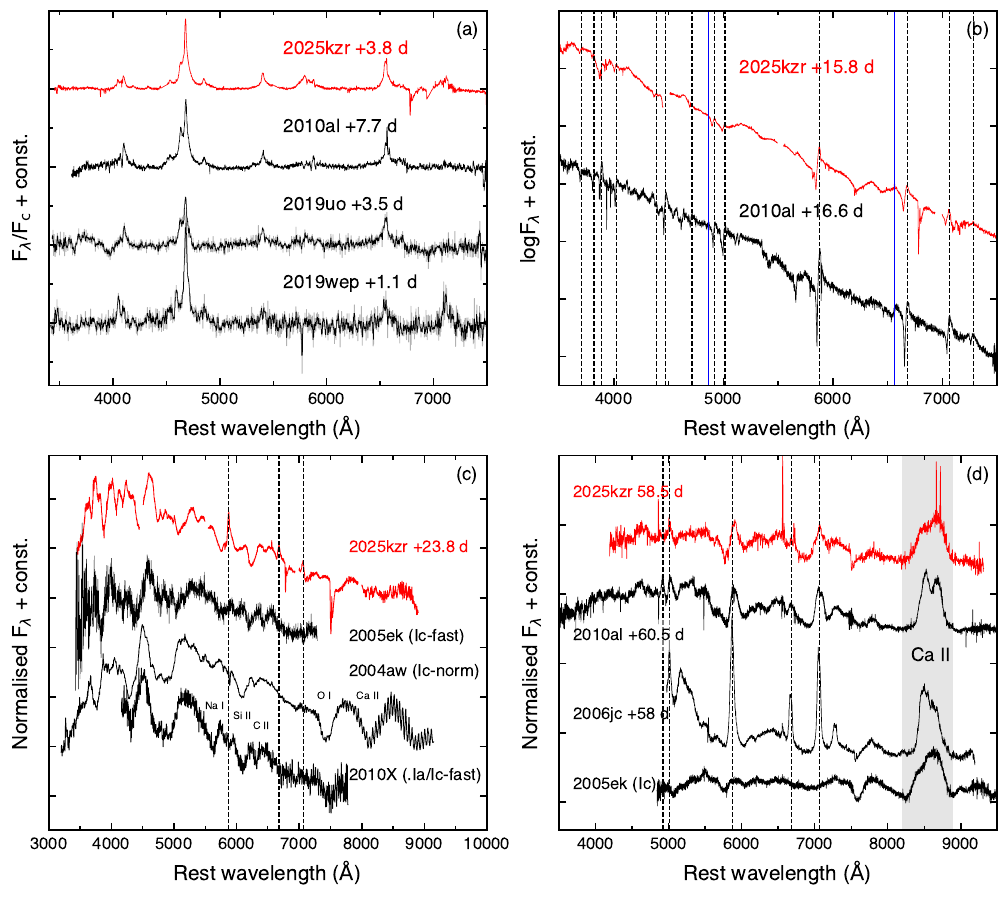}
    \caption{Spectral comparison during the flash (a), Helium P Cygni (b), ejecta (c), and late-time phases (d). The comparison SN~Ic spectra in panel (c) are all close to maximum light, and are shown due to their spectral similarity to our 23.8-day spectrum. The SN~2005ek spectrum in panel (d) is at 9 days post-maximum. Vertical dashed lines denote He lines, while the solid blue lines in panel (b) denote the H Balmer lines.  }
    \label{fig:spectra_comparison}
\end{figure*}

In Figure \ref{fig:spectra_comparison} we compare our SALT spectrum with three previous SNe Ibn showing flash features in their early spectra: SNe~2010al \citep{Pastorello2015}, 2019uo \citep{Gangopadhyay2020}, and 2019wep \citep{Gangopadhyay2022}. There are clear similarities between the objects in the sample, with all spectra showing prominent \ion{He}{ii}~$\lambda\lambda4686,5411$, \ion{N}{iii}~$\lambda4103$, and \ion{C}{iii}/\ion{N}{iii} $\lambda4650$ features. All spectra also show strong emission near H$\alpha$, which may in fact be \ion{He}{ii} based on our Pickering decrement analysis. Besides \SN{}, the highest-resolution spectrum belongs to SN~2010al. We have verified using the Pickering decrement method that the spectrum of SN~2010al is also H-free and we suggest that this may be the case for SNe 2019uo and 2019wep. The lower quality spectra for these objects precludes us from testing this rigorously. The wider implication is that the dense CSM in all these cases, and possibly for most SNe Ibn, is H-free \citep[although see SNe 2005la and 2011hw;][]{Pastorello2008, Smith2012, Pastorello2015}. 

Similar to \SN{} the spectrum of SN~2010al shows higher-order lines of the Pickering series, while the \ion{N}{iii}~$\lambda4103$ P Cygni feature is even more prominent than it is in \SN{}, as is the \ion{C}{iii}/\ion{N}{iii}~$\lambda4650$ blend. \ion{He}{i}~$5876$ is also visible. Interestingly, the very first spectrum\footnote{Unpublished by \citet{Pastorello2015} but available on the Transient Name Server} for SN~2010al from two days prior shows even stronger \ion{He}{i}~$5876$ in emission as well as \ion{He}{i}~$6678$, which is reminiscent of the strong \ion{He}{i} emission lines seen in our first Mookodi spectrum. Using the spectrum of SN~2010al presented in Figure \ref{fig:spectra_comparison} we have verified that the higher-ionisation lines appear to be blueshifted compared to the \ion{He}{i}~$5876$ line, which is at zero velocity when using the redshift of the narrow host H$\alpha$ line. This mirrors what we found for \SN{} and alludes to similar physical processes occurring in both systems. 

\subsubsection{Boian and Groh models}\label{subsubsec:BG}
We now consider the grid of model spectra presented in \citet{Boian2020}, as these are the only publicly-available models which include He-rich (although not H-free) surface abundances appropriate for \SN{}. The models were computed using the non-LTE radiative transfer code \verb|CMFGEN| as described in \citet{Groh2014} and \citet{Boian2019} and are meant to simulate the spectra of interacting SNe in the immediate days post-explosion. The hydrodynamics of the SN and CSM interaction are not considered, and instead a simplified setup is employed in which an energy source with a luminosity $L$ heats up the progenitor wind at an inner boundary $R_\mathrm{in}$. The energy source is widely believed to come from the CSM-shock interaction in which kinetic energy is converted to radiation. This energy source also powers the light curve. 

Besides $L$ and $R_\mathrm{in}$, the physical parameters which are input to the models include the progenitor mass-loss rate $\dot{M}$, the terminal wind velocity $v_\infty$, and the surface abundances. Three surface abundances are adopted: solar-like, CNO-processed and He-rich. The solar-like abundances correspond to low-mass RSG progenitors with massive H envelopes and little to no CNO-processed material at the stellar surface; the CNO-processed models correspond to massive RSGs, yellow hypergiants (YHGs), and blue supergiants (BSGs) with CNO-processed material at the surface due to rotational mixing, mass-loss, or binary interaction; and the He-rich models correspond to luminous blue variable (LBV) and late-type WN stars with low H abundances. The solar-like and CNO-processed models both have relative abundances by mass of 0.70 for H and 0.28 for He, but with a slightly higher N abundance in the CNO-processed case. By contrast, the He-rich models have abundances of 0.18 for H and 0.80 for He. The progenitor mass-loss rates range from $10^{-3}$ to $10^{-2}$~M$_\odot$~yr$^{-1}$, while the SN luminosities range from $1.9\times10^{8}$ to $2.5\times10^{10}$~L$_\odot$. Three radii are considered ($[8,16,32]\times10^{13}$~cm) which correspond to times of 1.0, 1.8, and 3.7~days post-explosion assuming a constant SN velocity of 10000~km~s$^{-1}$. 

\begin{figure}
    \centering
    \includegraphics[]{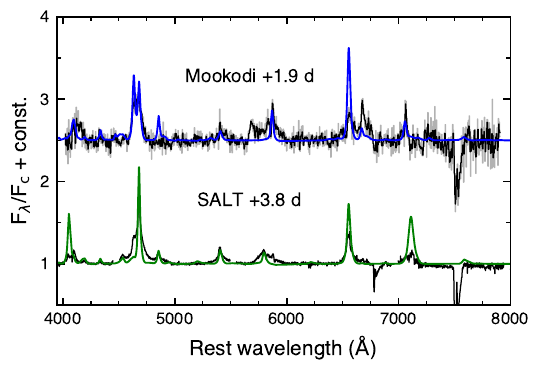}
    \caption{Our normalised Mookodi 1.9 day and SALT 3.8 day spectra with the best \citet{Boian2020} models. The Mookodi model (blue) has $L=7.8\times10^8$~L$_\odot$, $R=8\times10^{13}$~cm and $\dot{M}=3\times10^{-3}$~M$_\odot$, while the SALT model (green) has $L=2.5\times10^{10}$~L$_\odot$, $R=32\times10^{13}$~cm and $\dot{M}=10^{-2}$~M$_\odot$. The Mookodi spectrum has been smoothed using a Savitzky-Golay filter. Model spectra were convolved with a Gaussian kernel with $\Delta v=1000$~km~s$^{-1}$ ($R\approx300$) to match the observed line widths. }
    \label{fig:BG_models}
\end{figure}

A constant wind velocity of $v_\infty=150$~km~s$^{-1}$ is adopted for all models, resulting in emission lines much narrower than those we observe in \SN{}. To match the observed line widths, we convolve the model spectra with a Gaussian kernel of $\Delta v=1000$~km~s$^{-1}$ ($R\approx300$). We do not perform a formal fit to the 1.9-day Mookodi and 3.8-day SALT spectra. Instead, we determine by eye which of the continuum-normalised He-rich models best reproduce the observed spectral features. Figure \ref{fig:BG_models} shows our preferred models for each spectrum. The 1.9-day model has $L=7.8\times10^8$~L$_\odot$, $R=8\times10^{13}$~R$_\odot$ and $\dot{M}=3\times10^{-3}$~M$_\odot$, while the 3.8-day model has $L=2.5\times10^{10}$~L$_\odot$, $R=32\times10^{13}$~R$_\odot$ and $\dot{M}=10^{-2}$~M$_\odot$. The lower temperature of the 1.9-day model\footnote{Derived using the Stefan Boltzmann law.} accounts for its stronger \ion{He}{i} lines and \ion{C}{iii}/\ion{N}{iii}~$\lambda4650$ blend. The 3.8-day model reproduces the observed spectrum well overall but overestimates the flux of the \ion{N}{iv}~$\lambda\lambda 4058,7115$ and H Balmer lines. For H, this reflects the model's higher H abundance (0.18) compared with what we expect based on the Pickering decrement analysis (Figure \ref{fig:WR_comparison}). For N, the discrepancy may indicate that the actual N abundance is lower in \SN{}. In the 1.9-day model, the Balmer lines are stronger than observed, and the weakness or absence of H$\beta$ in the observed spectrum supports a H-free CSM. Importantly, in all of the models we considered the \ion{C}{iii}~$\lambda5696$ line is far weaker (or absent entirely) compared to what we observe in the 1.9-day spectrum, which may suggest a higher C abundance. 

To obtain actual mass-loss rates it is necessary to apply a number of scaling laws. As noted by \citet{Boian2020}, once a model which matches the observed lines is found, the luminosity usually has to be scaled to match the observed luminosity. The mass-loss rate must then be adjusted following the relation $\dot{M}\propto L^{3/4}$ from \citet{Grafener2016}. Different values of the terminal wind velocity will also change the mass-loss rate since the mass continuity equation ensures that $\dot{M}/v_\infty$ remains constant. Our 3.8-day model is only a factor of ${\sim}3$ fainter than the actual spectrum, whereas the Mookodi model underpredicts the flux by a factor of ${\sim}16$. Assuming a terminal wind velocity of ${\sim}1500$~km~s$^{-1}$ derived from the \ion{He}{i} lines (Section \ref{subsec:he_phase}) we find a mass-loss rate of ${\approx}0.2$~M$_\odot$~yr$^{-1}$ for both models, which should be regarded as a rough estimate. We caution that the models presented here are not the only ones consistent with the observations, and none show perfect agreement. Detailed modeling of the \SN{} flash spectra \citep[e.g.][]{Dessart2017,Dessart2023} is needed to reproduce the observed abundances and line profiles, which is beyond the scope of the present work.  

\subsection{Helium P Cygni phase}\label{subsec:he_phase}
By 11 days post-explosion the flash lines in \SN{} had completely disappeared. Thereafter, the spectrum was dominated by absorption lines of \ion{He}{i}, with the most prominent lines (e.g. \ion{He}{i}~$\lambda5876$) showing P Cygni profiles. In the standard physical picture for SNe IIn and Ibn SNe \citep{Smith2017HSN}, this spectral transformation is attributed to the photosphere receding into the shocked CSM where a cold dense shell \citep[CDS;][]{Chugai2004} formed at the contact discontinuity between the forward and reverse shocks is continuously reheated by X-ray and UV radiation and gives rise to intermediate-velocity (few $10^3$~km~s$^{-1}$) lines like H$\alpha$ and \ion{He}{i}. 

Our highest-resolution spectrum during this phase is our SALT spectrum at 15.8 days, which we compare with an X-Shooter spectrum of SN~2010al \citep{Pastorello2015} at a similar phase in Figure \ref{fig:spectra_comparison}. Both spectra show a large number of absorption lines of \ion{He}{i}. We identify at least 12 lines in the optical range in our SALT spectrum. The SN~2010al spectrum also shows weak H$\alpha$ in absorption, whose identity is confirmed via a later spectrum at 26 days which shows a similar P Cygni profile and blueshifted velocity (1300~km~s$^{-1}$) to the adjacent \ion{He}{i}~$\lambda6678$ line, as well as weaker H$\beta$ in absorption. 
Previous studies have shown that some SNe Ibn exhibit H in their spectra \citep[e.g., SNe 2005la, 2011hw;][]{Pastorello2008, Smith2012, Pastorello2015}, and it has been suggested that these objects may be transitioning from the LBV phase to a late-type WN Wolf-Rayet phase. In contrast, \SN{} does not show evidence for trace H, which may indicate a more advanced evolutionary state. 

\begin{figure}
    \centering
    \includegraphics[]{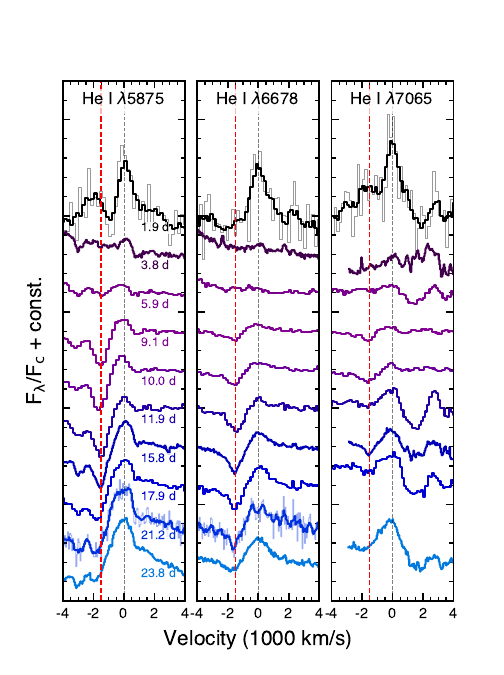}
    \caption{Evolution of the \ion{He}{i}~$\lambda\lambda5875$, $6678$, and $\lambda7065$ lines between 1.9 and 23.8 days post-explosion. Red vertical lines indicate a velocity of $-1500$~km~s$^{-1}$, which coincides with the P Cygni absorption minimum.}
    \label{fig:HeI_evolution}
\end{figure}

In Figure \ref{fig:HeI_evolution} we show the evolution of the \ion{He}{i}~$\lambda\lambda5876$, $6678$ and $\lambda7065$ lines between 1.9 and 23.8 days post-explosion. Our dense spectral sampling shows that the blueshifted absorption component of each line remains at a constant velocity of $-1500$~km~s$^{-1}$ from when it begins to appear at 5.9 days. At later times a higher-velocity absorption component develops (not shown), which is likely related to the SN ejecta and which we discuss in Section \ref{subsec:ejecta} below. A constant velocity P Cygni absorption minimum has been observed before in interacting SNe \citep[e.g. SN 1994W;][]{Sollerman1998,Chugai2004}. The interpretation of where the line forms and whether the velocity reflects the velocity of the CDS or unshocked CSM is still debated. We note, however, that a velocity of ${\sim}1500$~km~s$^{-1}$ is typical for WR wind speeds \citep{Crowther2007}, which may support the unshocked CSM scenario.

\begin{figure}
    \centering
    \includegraphics[]{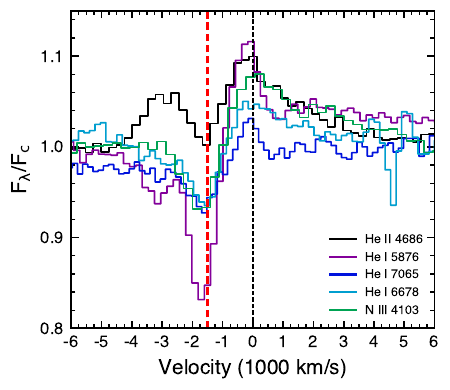}
    \caption{Line profiles for \ion{He}{ii}~$\lambda4686$, \ion{N}{iii}~$\lambda4103$, \ion{He}{i}~$\lambda5875$, $\lambda6678$, and $\lambda7065$ from our NTT 10.0 day spectrum. The red vertical line indicates a velocity of $-1500$~km~s$^{-1}$.}
    \label{fig:NIII_line_profile}
\end{figure}

One final noteworthy spectral feature during this Helium P Cygni phase is the \ion{N}{iii}~$\lambda4103$ line. This feature was regarded as H$\delta$ by \citet{Pastorello2015b} in the first spectrum of SN~2010al but it is more likely to be the \ion{N}{iii} $\lambda\lambda4097,4103$ doublet seen in the spectra of hot Wolf-Rayet and O-type stars. The feature has a pronounced P Cygni profile very similar to that seen in the \ion{He}{i} lines in the same spectrum at 10.0 days (Figure \ref{fig:NIII_line_profile}). In addition, the \ion{He}{ii}~$\lambda4686$ line also appears to have a similar profile. The \ion{N}{iii}~$\lambda4103$ line was visible during the flash phase but with no absorption component, which might be due to contamination from the nearby \ion{N}{iv} line or optical depth effects. It also appears to have been visible in emission in the flash spectra of other SNe Ibn (Figure \ref{fig:spectra_comparison}). By 12 days, however, the prominent P Cygni feature had disappeared entirely along with the other higher-ionisation lines. The physics of \ion{N}{iii} line formation is highly complex and involves competing non-LTE processes such as dielectronic recombination and the wind-driven Swings mechanism, as discussed in the seminal work by \citet{Mihalas1973} and the more recent work of \citet{RiveroGonzalez2011}, both of which focus on \ion{N}{iii} line formation in O-type stars. 

\subsection{Ejecta phase}\label{subsec:ejecta}
Already in our 15.8-day spectrum (Figure \ref{fig:spectra_comparison}) we begin to see broader features developing, the most prominent being the asymmetric absorption dip at ${\sim}6200$~\AA~and a shallower dip at ${\sim}5300$~\AA. By 23.8 days these features have developed into even broader absorption features more commonly seen in photospheric spectra of Type Ibc SNe (Figure \ref{fig:spectra_comparison}). This suggests that the continuum photosphere has receded and we are now seeing the SN ejecta.  

Using the SNID-SAGE spectral matching tool \citep{Stoppa2026}, we find that our SALT 23.8-day spectrum is most closely matched with Type Ic SNe, with the top matches being SNe~2005ek and 2004aw \citep{Drout2013,Taubenberger2006}. SN~2005ek, studied in detail by \citet{Drout2013}, is notable as one of the fastest-evolving SNe Ic, spectroscopically similar to normal SNe Ic but with accelerated evolution. Similar to their Figure 9, in Figure \ref{fig:spectra_comparison} we compare our 23.8 day spectrum with spectra of SNe 2005ek, 2004aw, and 2010X \citep{Kasliwal2010} near their maximum light. Besides the \ion{He}{i}~$\lambda\lambda5875$, $6678$, and $\lambda7065$ lines, which are still clearly seen in emission, the spectrum of \SN{} bears a strong resemblance to the comparison spectra. Even when excluding the \ion{He}{i} emission features when spectral template matching, SNID SAGE finds the same Type Ic matches. The phase relative to peak brightness, however, is much later in the case of \SN{} at 12.7 days past the $r$-band peak. As demonstrated in Section \ref{subsec:modelling}, however, the main peak in \SN{} is likely powered by CSM-ejecta interaction, whereas the comparison objects were all shown to be powered by radioactivity. This may explain the discrepancy in phase. 

The maximum-light spectra for SN~2005ek were modelled by \citet{Drout2013} with a combination of \ion{O}{i}, \ion{C}{ii}, \ion{Mg}{ii}, \ion{Si}{ii}, \ion{Ca}{ii}, \ion{Ti}{ii}, and \ion{Fe}{ii} at a velocity of 8000--9000~km~s$^{-1}$. We highlight some of these features in Figure \ref{fig:spectra_comparison}. If we assume the prominent absorption feature at ${\sim}6200$~$\AA$ in our spectra is also \ion{Si}{ii}, we estimate a slower photospheric velocity of ${\sim}5000$~km~s$^{-1}$ for \SN{}. This is slower than velocities observed in the comparison spectra, and may signify a later phase past the radioactive-powered peak, or, if observed at peak, a lower kinetic energy or higher ejecta mass involved in the explosion. 

How confident are we in the Type Ic classification for our 23.8 day spectrum, given the contamination from the \ion{He}{i} lines produced via CSM interaction? It is possible that the absorption blueward of \ion{He}{i}~$\lambda5875$ is blueshifted \ion{He}{i} rather than \ion{Na}{i}, which would make the spectrum more similar to that of a SN~Ib rather than a SN~Ic. We do not, however, unambiguously detect other \ion{He}{i} lines in absorption that one would expect for SNe Ib, but this might be simply a result of contamination from the \ion{He}{i} emission lines. One means of separating Ib and Ic SNe is to compare the relative depths of the $\lambda6200$ and \ion{O}{i}~$\lambda7774$ absorption troughs in their peak spectra \citep{Matheson2001,Galyam2017HSN}. SNe Ib tend to have shallower \ion{O}{i}~$\lambda7774$ absorption relative to $\lambda6200$, whereas the opposite is generally true for SNe Ic. We measure a relative depth ratio of unity for \SN{}, which is still ambiguous. Additionally, \citet{Fremling2018} and \citet{Shivvers2019} showed that the pEW of \ion{O}{i}~$\lambda7774$ line is generally much larger for SNe Ic than for SNe Ib and IIb; the pEW we measure for the \ion{O}{i}~$\lambda7774$ is more consistent with the Ib samples presented in these works. Overall, although our 23.8 day spectrum suggests a Ic-like ejecta we emphasise that there is still a degree of uncertainty.   

\subsection{Late-time phase}\label{subsec:nebular}
The final spectrum of \SN{} (Figure \ref{fig:spectra_comparison}) was obtained at 58.5 days post-explosion, just prior to solar conjunction. The most prominent spectral features are the \ion{Ca}{ii} near infrared (NIR) triplet in emission as well as the \ion{He}{i}~$\lambda7065$ and $\lambda5875$ lines. The emission components of the \ion{He}{i} lines are broader than during the Helium P Cygni phase (see the increase in EW in Figure \ref{fig:EW_evolution}), and we also identify a blueshifted absorption component at a velocity of $\approx5000$~km~s$^{-1}$. Figure \ref{fig:spectra_comparison} demonstrates that the late-time spectrum of \SN{} closely resembles that of SN~2010al at the same phase, sharing a similar broad P Cygni \ion{He}{i} morphology and a similar continuum shape. For comparison we also show a spectrum of the prototypical SN~Ibn 2006jc at the same phase, as well as SN~Ic 2005ek which we show due to its spectral similarity during the ejecta phase. The spectra of SN~2006jc and SN~2005ek illustrate important differences with respect to those of SNe~2010al and 2025kzr. \citet{Hosseinzadeh2017} proposed two spectral classes for SNe~Ibn: those with P Cygni \ion{He}{i} lines and those in which \ion{He}{i} is mostly in emission. SN~2006jc falls into the latter subclass due to its strong and narrow \ion{He}{i} emission lines, whereas SNe~2010al and 2025kzr belong to the former class due to their broader, P Cygni-like \ion{He}{i} profiles. Another difference is a lack of the strong blue pseudo-continuum which is clearly seen rising towards the blue in the SN~2006jc spectrum. Figure \ref{fig:spectra_comparison} demonstrates that even though \SN{} resembled the Type~Ic SN~2005ek during the earlier ejecta-dominated phase, at later times it is typical of SN~Ibn (of the P Cygni class) rather than a SN~Ic.

\section{Polarimetric properties}
\begin{figure}
    \centering
    \includegraphics[]{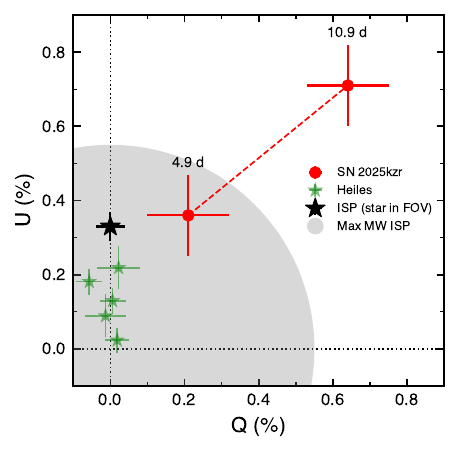}
    \caption{Stokes $Q-U$ plane showing $V$-band imaging polarimetry results for \SN{}. The first epoch at 4.9~d is within the maximum level of the MW ISP (grey circle). For comparison we show polarisation measurements for stars within 5~deg of \SN{} in the \citet{Heiles2000} catalogue (green), as well as the ISP measured from a star in the FOV (black). Using this star to correct for Galactic ISP, we measure an intrinsic polarisation of $P=0.21\pm0.12 \%$ and $P=0.74\pm0.12\%$ for the first and second epochs, respectively.}
    \label{fig:pol}
\end{figure}

Figure \ref{fig:pol} presents the results of our polarimetric observations. For both epochs at 4.9 and 10.9 days, \SN{} is offset from the comparison stars and therefore shows evidence for intrinsic polarisation, although we note that the first epoch is still consistent with Galactic ISP. The first epoch polarisation degree of $P=0.21\pm0.12\%$ suggests a high degree of spherical symmetry, and implies that the flash-ionised CSM surrounding the progenitor is spherical. If the precursor outburst at 55 days before explosion created a spherical CSM this may point towards a single-star rather than binary-driven mass loss event (see Section \ref{subsec:precursor} below). Our second epoch at 10.9 days shows a significantly higher polarisation degree at $P=0.74\pm0.12\%$ which signals a deviation away from spherical symmetry. At this time the flash-ionisation lines had completely disappeared and the spectra were dominated by \ion{He}{i} P Cygni lines. Five days later the spectrum was beginning to show ejecta features, and at 23.8 days the spectrum was reminiscent of (certain) SN Ic spectra. The increasing polarisation degree might imply an aspherical explosion, and would be consistent with observations of SNe Ic which are known to show significant polarisation \citep{Wang2008}. 

Polarimetric observations of SNe Ibn/Icn are rare, with known cases including SNe~2015G \citep{Shivvers2017}, 2015U \citep{Shivvers2016}, 2021csp \citep{Perley2022}, 2023emq \citep{Pursiainen2023}, and 2024abvb \citep{Intel2026}. Spectropolarimetry of SN Ibn 2015G showed $P\sim2.7\%$ at 5 days post-discovery, indicating a high degree of asymmetry, whereas the Type Icn SN 2021csp and transitional Type Ibn/Icn SN~2023emq both showed polarisation degrees consistent with spherical symmetry. Similar to \SN{}, SN~2024abvb showed low-to-high transition in the polarisation degrees after peak brightness, although the increasing polarisation was attributed to dust rather than an aspherical Ic ejecta. Differences in the geometry of the CSM for these objects is not unexpected given that the progenitor channels to SNe Ibn/Icn are likely diverse. Further polarimetric observations of H-poor interacting SNe are needed to shed light on the CSM geometry. 

\section{X-ray and radio observations}\label{sec:X_radio}
The ejecta-CSM interaction shock is expected to generate nonthermal radio synchrotron emission and free-free X-ray emission \citep{Chevalier1982,Chevalier2017HSN,Inoue2025}. Observations in these regimes can provide powerful constraints on the CSM properties. Here we consider our X-ray and radio non-detections.  

\begin{figure*}
    \centering
    \includegraphics[]{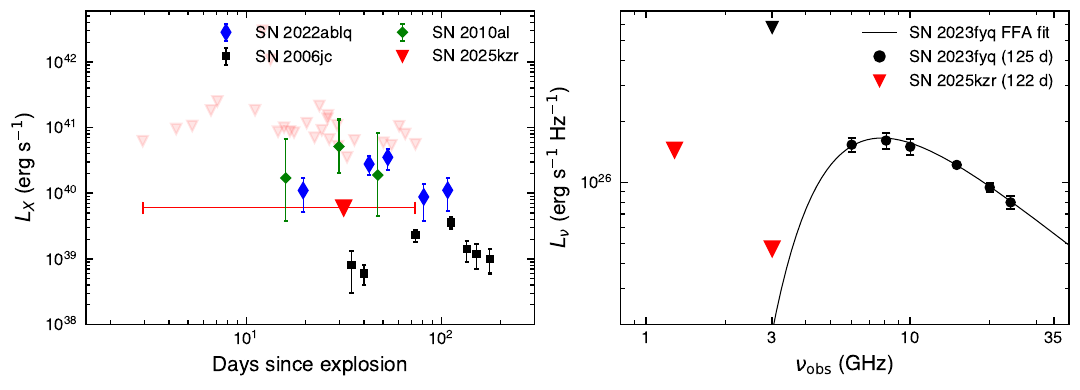}
    \caption{Left: X-ray upper limits for \SN{} compared to the three previous SNe Ibn with X-ray detections: SNe~2006jc \citep{Immler2008}, 2010al \citep{Ofek2013b}, and 2022ablq \citep{Pellegrino2024}. The faded red upper limits denote single-observation upper limits, while the solid red upper limit was derived from combining all of our X-ray observations. The x-error bars indicate the timespan over which \Swift{}/XRT observations were obtained. Right: radio upper limits for \SN{} compared to the SED for SN~2023fyq at a similar epoch. The solid line indicates the best-fit FFA model from Figure 2 in \citet{Baerway2025}.  }
    \label{fig:X_radio}
\end{figure*}

Only three SNe Ibn have been detected in X-rays: SNe 2006jc \citep{Immler2008}, 2010al \citep{Ofek2013b}, and 2022ablq \citep{Pellegrino2024}. These studies found elevated mass loss ${\sim}0.5-2$~yrs before explosion that agrees with conclusions drawn from optical observations. \Swift{} did not detect X-ray emission from \SN{}. Figure \ref{fig:X_radio} demonstrates that our single-epoch upper limits would not have been deep enough to detect SNe~2006jc and 2022ablq, and would have been on the edge of detectability for SN~2010al. The upper limit from combining all of our X-ray observations would have been deep enough to detect emission similar to that of SNe~2010al and 2022ablq but not SN~2006jc. 

\citet{Inoue2025} recently developed a model for the X-ray light curves of SNe Ibn/Icn and applied it to the soft X-ray light curves of SNe Ibn 2006jc and 2022ablq. They also considered the SN Icn~2019hgp, which was not detected by \Swift{}. In their model, the hard X-rays (10--40~keV) can robustly constrain the CSM density, explosion energy, and ejecta mass, whereas the soft X-rays (0.2--10~keV) encode information about the CSM composition because the rising phase is controlled by photoelectric absorption in the unshocked CSM. An important prediction of their model is a double-peaked soft X-ray lightcurve in which the early bright emission is caused by complete photoionisation of the unshocked CSM; as the luminosity drops and the CSM recombines the optical depth rises and there is a sharp dip before rising again to the second peak. Similar to SN~2019hgp, our early non-detections rule out the predicted early peak and imply either an interaction shock that is too weak to fully ionise the unshocked CSM ($\xi\gtrsim100$ in the \citet{Inoue2025} framework) or a CSM dense enough for Compton scattering alone to attenuate the early flash. 

According to \citet{Inoue2025}, the lower luminosity and later peak time for SN~2006jc compared to SN~2022ablq (Figure \ref{fig:X_radio}) are explained by a more metal-rich CSM composition (with $(\mathrm{He,C,O})=(0.4,0.3,0.3)$ for SN~2006jc compared to $(0.95,0.025,0.025)$ for SN~2022ablq). This is because the photoelectric opacity has a strong dependence on metallicity \citep[$\kappa_{\mathrm{pe,}j}\propto Z_j^5$ for the K-shell approximation;][]{Longair2011}. A progenitor more strongly stripped of its He envelope could therefore explain the absence of X-ray emission in \SN{}, particularly in light of our very first spectrum showing \ion{C}{iii} lines more commonly seen in SNe Icn. We note, however, that our X-ray upper limits are not deep enough to place stringent constraints on these models, so we refrain from pursuing further modelling.   

Recently, \citet{Baerway2025} reported the first radio detection of a SN Ibn. They found that the SED of SN~2023fyq was best fitted by a synchrotron spectrum attenuated by free-free absorption (FFA) from the CSM, with an inferred mass-loss rate of ${\sim}4\times10^{-3}~M_\odot~$yr$^{-1}$ from 0.7 to 3 yr before explosion, consistent with constraints from optical observations. We obtained a single epoch of radio observations at 122~days with MeerKAT at 1.28 and 3~GHz, both of which yielded upper limits. In Figure \ref{fig:X_radio} we show these limits alongside the SED of SN~2023fyq and the best-fit FFA model from \citet{Baerway2025}. Although our upper limits are consistent with this model, without higher frequency observations we cannot determine whether the low-frequency flux is being suppressed by FFA or synchrotron self absorption. If FFA is dominant, as in SN~2023fyq, our upper limits imply a similar CSM density and mass-loss rate. 

\section{Discussion}\label{sec:discussion}

\subsection{Blueshifted higher-ionisation lines}\label{subsec:blueshift}
One of the most striking findings from our 3.8 day spectrum was that the higher-ionisation flash lines (e.g. \ion{He}{ii}) are blueshifted by 460~km~s$^{-1}$ compared to the \ion{He}{i} emission lines (Figure \ref{fig:SALT_line_profiles}). The reason we are able to discern such a blueshift in our low-resolution spectra is because the CSM velocities observed in SNe Ibn (${\gtrsim}1000$~km~s$^{-1}$) are higher than in the RSG progenitors of SNe IIP (${\sim}35$~km~s$^{-1}$). 
Blueshifted flash-ionised emission lines have only been observed in a handful of cases \citep[SNe 1998S and 2023ixf;][]{Shivvers2015,Smith2023}. According to \citet{Smith2023} this can be understood as a combination of light travel time effects and occultation of the far side of the CSM by the photosphere \citep{Groh2014,Shivvers2015,Grafener2016}. The fact that our higher-ionisation lines are blueshifted while the lower-ionisation lines are not can be understood in the framework outlined by \citet{Shivvers2015}. The ionisation structure in the progenitor wind is such that the higher-ionisation lines (e.g. \ion{He}{ii}, \ion{C}{iv}) form at smaller radii compared to lower-ionisation lines such as \ion{He}{i}. From the perspective of a point on the far (receding) side of the SN, the solid angle subtended by the photosphere is larger at smaller radii, so emission from inner regions is preferentially occulted compared to outer regions. Higher-ionisation lines therefore have a greater deficit in redshifted emission, and their lines are consequently blueshifted compared to the systemic velocity. This is our preferred interpretation. Other works have shown that blueshifted emission lines arise naturally in an expanding outflow, with a stronger blueshift as a function of CSM velocity \citep{Grafener2016,Huang2018}. However, in this interpretation we would expect all lines (including lower-ionisation lines) to be blueshifted in a similar fashion, and more complex scenarios would be needed (e.g. a steep negative velocity gradient) to explain our observations. 

How do we explain the diminishing blueshift of the higher-ionisation lines? As the luminosity and photospheric temperature decrease, the flash lines decrease in strength (Figure \ref{fig:EW_evolution}), and by 10 days the line profiles and line centres of both higher and lower-ionisation lines begin to look similar (Figure \ref{fig:NIII_line_profile}). This can be explained by the fact that as the forward shock sweeps through the unshocked CSM, the inner radii, where the higher-ionisation lines preferentially form, are overrun first and contribute progressively less emission to the line profiles. The occultation effect diminishes since both higher and lower-ionisation lines form at similar radii. More complex scenarios such as radiative acceleration \citep[e.g.,][]{Chugai2002,Fransson2014} or an asymmetric CSM with a disk or torus geometry \citep[e.g.][]{Smith2023} can be invoked to explain our observations and might also be consistent with the increasing deviation from spherical symmetry implied by our polarimetric observations, but we choose not to consider these further. 

Regarding the velocity of the unshocked CSM, we measured a FWHM of 1200~km~s$^{-1}$ for the narrow component of the \ion{He}{ii} lines (Figure \ref{fig:SALT_line_fits}). Because light from the far (receding) side is missing, determining the actual CSM speed is not straightforward \citep[as noted by][for SN~2023ixf]{Smith2023}, although we note that the blue edge of the \ion{He}{ii} lines is the same as the velocity of $1500$~km~s$^{-1}$ measured from \ion{He}{i} P Cygni absorption at later epochs (Figure \ref{fig:HeI_evolution}). We emphasise that more detailed modelling of the blueshifted lines is needed to determine the actual outflow speed and various other properties (optical depth, temperature and ratio of inner to outer radius) of the CSM.

\subsection{Constraining the properties of the CSM}
\subsubsection{Disappearance of flash lines and connection to precursor}\label{subsubsec:timing}
One of the key observed timescales in \SN{} is the disappearance of the flash-ionised emission lines at ${\sim}10$ days post-explosion. Combined with the CSM expansion speed measured from the \ion{He}{i} P Cygni profiles we can use this to estimate the time of the putative pre-SN mass-loss event. Assuming a typical shock velocity of $10^4$~km~s$^{-1}$, the forward shock would have reached a radius $R=v\Delta t=8.6\times10^{14}$~cm by 10~days which corresponds to the outer edge of the dense CSM. For a CSM speed of 1500~km~s$^{-1}$, a pre-SN mass ejection would have occurred at $t_\mathrm{pre-SN}=R/v_\mathrm{CSM}=t_\mathrm{obs}\times(v_\mathrm{sh}/v_\mathrm{CSM})\approx 66$~days before explosion. This is close to the observed onset of the precursor outburst at 55~days and strongly implies that the precursor was responsible for forming the dense CSM.

\subsubsection{Shock breakout in a dense wind}
The first few days of observations suggest that the radiation from the SN shock was breaking out of a region of dense CSM ejected during the precursor outburst. This is supported by the fast rise to peak bolometric luminosity at $t_\mathrm{p}=5.8$~days, UV-bright emission with temperatures $T{\geq}25000$~K, flash-ionised emission lines that persist until ${\sim}10$~days, and intermediate-velocity ($v\approx1500$~km~s$^{-1}$)~\ion{He}{i} P Cygni lines that follow the flash phase. 

We estimate the CSM mass and mass-loss rate using two approaches. First we use a model-independent estimate for the characteristic luminosity associated with shock breakout in an extended CSM \citep{Waxman2017,Zimmerman2024},
\begin{equation}
L_\mathrm{bo}\approx \frac{M_\mathrm{CSM}v_\mathrm{bo}^2}{t_\mathrm{bo}}. 
\end{equation}
Using the peak luminosity of ${\approx}10^{44}$~erg~s$^{-1}$ from our bolometric light curve, and a breakout time equal to the time of peak luminosity (5.8~days), we estimate a CSM mass of ${\gtrsim}0.03~M_\odot$. Here we assumed a breakout velocity of $10^4$~km~s$^{-1}$. This mass estimate is likely a lower limit on the total CSM mass since it represents only the shocked CSM mass at breakout. The corresponding mass-loss rate is ${\gtrsim}0.2~M_\odot$~yr$^{-1}$ if we assume the mass loss started at the time of the observed precursor, 55 days before explosion. This value is consistent with the mass-loss rate derived from our flash spectra (Section \ref{subsubsec:BG}).  

The second approach follows \citet{Chevalier2011}. After explosion, a radiation-dominated shock moves through the freely-expanding SN ejecta and encounters the CSM, where a shocked layer bounded by forward and reverse shocks develops. For the shock to remain radiation-dominated, the characteristic optical depth $\tau_\mathrm{sh}{\approx}c/v_\mathrm{sh}$ must be smaller than the optical depth within the wind, $\tau_\mathrm{w}>\tau_\mathrm{sh}$. Radiation can escape the system when the diffusion time equals the expansion time. For the case of a steady wind with mass-loss rate $\dot{M}$ and wind velocity $v_\mathrm{w}$, the density is 
\begin{equation}
\rho_\mathrm{w}=\frac{\dot{M}}{4\pi r^2 v_\mathrm{w}}\equiv D r^{-2},
\end{equation}
 where $D$ is a constant that can be recast in terms of a dimensionless parameter $D_\star$ scaled to a mass-loss rate of $10^{-2}~M_{\odot}$~yr$^{-1}$ and a wind velocity of $10$~km~s$^{-1}$, so that $\rho_\mathrm{w}=5.0\times10^{16}D_\star r^{-2}$ in cgs units. \citet{Chevalier2011} define the diffusion radius $R_\mathrm{d} \equiv \kappa D v_\mathrm{sh}/c$ by the breakout condition $\tau_\mathrm{w}\approx c/v_\mathrm{sh}$, and consider the two scenarios with $R_\mathrm{d}<R_\mathrm{w}$ and $R_\mathrm{d}>R_\mathrm{w}$. For \SN{} we assume that shock breakout occurs within the dense CSM ($R_\mathrm{d}<R_\mathrm{w}$), similar to \citet{Margutti2014} and \citet{Terreran2022} for SNe~2009ip and 2020pni, respectively. This is supported by the disappearance of the flash lines at ${\sim}10$~days post-explosion which occurs well after the time of peak luminosity, and indicates that the shock had not yet reached the edge of the dense CSM. \citet{Chevalier2011} showed that the diffusion time does not depend on the radius, but only the opacity and density of the CSM
\begin{equation}
t_\mathrm{d}=\frac{R_\mathrm{d}}{v_\mathrm{sh}}=\frac{\kappa D}{c}=6.6 k D_\star ~\mathrm{days},
\end{equation}
where $k$ is the opacity in units of 0.34~cm$^2$~g$^{-1}$. For \SN{} we adopt an opacity of $\kappa=0.1$~cm$^2$~g$^{-1}$ appropriate for a He-rich composition. If we assume the rise time to peak luminosity is the time of shock breakout of the dense CSM, $t_\mathrm{rise}\approx t_\mathrm{bo}\approx t_\mathrm{d}$, we can estimate the mass-loss rate and mass of the dense CSM. Adopting as our rise time the time of peak luminosity (5.8~days, Figure \ref{fig:Lbol}) and the CSM velocity from the \ion{He}{i} lines (1500~km~s$^{-1}$, Figure \ref{fig:HeI_evolution}), we estimate a mass-loss rate of $\dot{M}\approx4.5~M_\odot$~yr$^{-1}$. If the mass ejection begins 55 days before explosion, consistent with the precursor, the total mass ejected is $M_\mathrm{CSM}=0.66~M_\odot$. The factor of ${\approx}20$ discrepancy between the two mass-loss rate estimates may point towards a CSM with a substantially different density profile than a normal stellar wind. Both mass-loss rates are high ($\dot{M}\geq10^{-1}$) and are inconsitent with the mass-loss rates observed in massive stars \citep{Smith2017HSN}, and clearly point towards a strong binary-driven or eruptive mass-loss episode.

\subsection{The precursor}\label{subsec:precursor}
Precursor emission is fairly common among SNe IIn \citep{Ofek2014,Strotjohann2021} but far rarer for other SN sub-types, with only a single Type IIP \citep[SN 2020tlf;][]{JacobsonGalan2022} and three SNe~Ibn (prior to \SN{}) showing precursor emission. We now discuss the mechanisms that can produce such emission. 

During the final stages of massive-star evolution, carbon or heavier elements are being fused in the core, which is predominantly cooled by thermal neutrinos \citep{Woosley2002}. Although the nuclear luminosity and neutrino cooling rates are largely in equilibrium, local convective instabilities arise due to the different temperature dependences of the nuclear burning and neutrino emission rates \citep{Quataert2012}. A significant fraction (${\sim}10\%$) of the fusion energy is carried away through convection, which can excite internal gravity waves \citep{Shiode2014}. For a subset of massive stars these waves may tunnel out of the core and excite acoustic waves that drive convection in the stellar envelope. The deposited energy of $10^{47}$--$10^{48}$~erg results in a super-Eddington power (${\sim}10^{41}$~erg~s$^{-1}$) that may trigger strong mass-loss events, potentially unbinding up to $1~M_\odot$ of material \citep{Quataert2012,Shiode2014,Fuller2017,Fuller2018}.

In the models of \citet{Shiode2014}, wave-driven mass loss can occur as early as ${\sim}10$~years before core collapse with the onset of neon fusion; oxygen fusion can drive mass loss on weeks to months timescales, and silicon fusion only in the final hours to days. The timing of the mass loss is inversely related to the core He mass, with heavier cores producing outbursts closer to explosion. Given that the precursor in \SN{} occurred two months before core collapse, we can set a rough upper limit on the He core mass of ${\lesssim}15~M_\odot$. Inferring the ZAMS mass, however, is not straightforward, as this depends on mixing and mass loss which both depend on rotation and metallicity \citep{Shiode2014}. The timing of the precursors to SNe 2006jc and 2019uo imply He cores with masses ${\lesssim}10~M_\odot$. 

\citet{Fuller2018} studied wave-driven mass loss in H-free stars and found characteristics broadly consistent with SNe~Ibn: mass-loss rates of ${\sim}0.1~M_\odot$~yr$^{-1}$, wind speeds of ${\sim}500$~km~s$^{-1}$ (at the lower end of observed \ion{He}{i} velocities), and CSM confined to ${\lesssim}10^{16}$~cm. They also suggested that shell-shell collisions could account for bright precursor emission in which ${\sim}10^{47}$~erg of kinetic energy is converted to thermal energy, analogous to the mechanism responsible for interaction-powered light curves. \citet{Wu2022} extended this work to ZAMS masses of 11--50~$M_\odot$ and found that for typical SN progenitors ($M_\mathrm{ZAMS}<30~M_\odot$), the wave energies excited during oxygen or neon burning are an order of magnitude lower than earlier estimates (${\sim}10^{46}$--$10^{47}$~erg between 0.1 and 10 yr before core collapse), too low to produce a detectable precursor. An important exception are the highest-mass progenitors ($M_\mathrm{ZAMS}\geq 30~M_\odot$) in which convective shell mergers boost the energy transmitted to the envelope by an order magnitude ($10^{47}$--$10^{48}$~erg) at ${\sim}0.01$--0.1~yr before core collapse. \SN{} radiated ${\approx}2\times10^{47}$~erg starting at 0.15~yr before explosion, which compares favourably with the high-mass predictions of \citet{Wu2022}, although a high radiative efficiency would be required. 

\citet{Tsuna2024} showed that it is difficult to obtain precursors with luminosities substantially brighter than the Eddington luminosity for pre-explosion stars. For example, a $10~M_\odot$ star has $L_\mathrm{Edd}{\approx}10^{39}$~erg~s$^{-1}$, which is fainter than most observed precursors \citep[$10^{40}$--$10^{42}$~erg~s$^{-1}$;][]{Strotjohann2021}. The peak $r$-band luminosity of the \SN{} precursor, ${\approx}10^{41}$~erg~s$^{-1}$, lies in this range. \citet{Tsuna2024} proposed two ways to break this luminosity limit. The first is to invoke the collision of two CSM shells, which raises the conversion efficiency of kinetic energy into radiation. The high-mass progenitor models mentioned above, transmitting ${\sim}10^{48}$~erg to the stellar envelope through wave heating, are promising channels for producing such shells. 

The second way to break the luminosity limit is to invoke super-Eddington accretion onto a compact object (CO) binary companion, as invoked for SN~2023fyq \citep{Dong2024}. In this model, a massive star in a binary undergoes a partial (or complete) envelope ejection that forms an homologously expanding CSM. Accretion onto the CO leads to an accretion disk that injects a disk wind back into the CSM. The kinetic energy of the wind is thermalised in the CSM and bright precursor emission is produced. \citet{Tsuna2024} considered three progenitors---a RSG, a high-mass compact He star, and a low-mass extended He star---and found that binary separations of 3--30 times the stellar radii can account for the luminosities and durations of observed precursors. Directly relevant to SNe~Ibn are the He star models. They found that the precursors to SNe 2006jc and 2019uo could be explained by a compact He star with a black hole (BH) companion and an inflated He star with a neutron star (NS) companion, respectively, suggesting diversity in the progenitor channels to SNe Ibn. The three-day duration and ${\approx}10^{41}$~erg~s$^{-1}$ luminosity of the \SN{} precursor are similar to those of SN~2019uo and might point towards a similar progenitor \citep[see Figure 4 in][]{Tsuna2024}, although several models remain roughly consistent with the data. 

In summary, the \SN{} precursor is plausibly linked to the late stages of nuclear burning. Single-star wave-driven mass loss in high-mass progenitors ($M_\mathrm{ZAMS}\geq 30~M_\odot$) can reproduce both the luminosity and the timescale, with the observed radiation arising from collisions between the ejected material and pre-existing CSM. A binary system comprising a He star and a CO can also reproduce these properties. In both cases, a phase of strong, eruptive mass loss in the final months before explosion is required. Importantly, the fact that the \SN{} precursor occurs much closer to explosion than those of SNe~2006jc and 2019uo (55 days versus ${\sim}1$ and ${\sim}2$~yr) strongly suggests a connection to core instabilities during the oxygen-burning phase.

\subsection{The progenitor} \label{subsec:progenitor}

The two most popular progenitor channels for SNe Ibn are (i) single massive WR stars exploding in a He-rich CSM ejected by the star itself, and (ii) lower-mass He stars in interacting binary systems where the CSM is formed through mass transfer or common-envelope evolution  \citep{Foley2007,Pastorello2008,Tominaga2008,Hosseinzadeh2017,Tsuna2024,Dong2024}. Recently, a direct progenitor detection to SN~2023fyq has been claimed \citep{Hong2026}, with a low-mass helium star in a binary system favoured over a single massive star, consistent with the findings of \citep{Dong2024}. 
 
One of the main arguments in favour of massive WR stars is the lower \textsuperscript{56}Ni masses inferred for SNe Ibn compared to the broader SESN population \citep{Maeda2022}. Most SESN progenitors share a similar ZAMS mass range to SNe IIP \citep[${\sim}8$--18$~M_\odot$;][]{Lyman2016}, with the difference between the two populations attributed to a strong binary interaction for SESNe \citep{Fang2019}. \citet{Maeda2022} argued that progenitors with higher initial masses (${\gtrsim18~M_\odot}$) can cause the lower \textsuperscript{56}Ni masses in SNe Ibn because they have a different Fe core structure---which influences the \textsuperscript{56}Ni production---compared to SESNe and SNe IIP. They provide an example of a $M_\mathrm{ZAMS}{\sim}25~M_\odot$ that will form an $8~M_\odot$ He star or $6~M_\odot$ C+O star that will eject 4.6--6.6~$M_\odot$ of material if a NS is formed and even less ejecta if a BH is formed. An explosion with the binding energy of the core (${\sim}10^{51}$~erg) would lead to substantial fallback onto the compact object, causing little to no \textsuperscript{56}Ni to be ejected \citep{Woosley1995,Maeda2007}. Furthermore, \citet{Maeda2022} hypothesised that the rarer class of SNe Icn results from even more massive progenitors than SNe Ibn and that there is a transition mass above/below which the final star is a C+O/He star. Due to the rate of SNe Icn being one tenth the rate of SNe Ibn they suggested a transition mass of $M_\mathrm{ZAMS}{\sim}40$--50$~M_\odot$. The `transitional' SNe Ibn/Icn such as SNe 2023emq and 2023xgo \citep{Pursiainen2023,Gangopadhyay2025} might be objects whose progenitors---if single massive stars---are close to this transition mass. 

Three objects which appear to rule out a massive star origin are PS1-12sk, SN~2024abvb, and SN~2024acyl. PS1-12sk exploded in a host showing no detectable star formation and may instead be related to a white dwarf binary system \citep{Sanders2013,Hosseinzadeh2019}, while SNe~2024abvb and 2024acyl were both found at large offsets from their host galaxies \citep{Cai2026,Dong2025,Hu2026,Intel2026,Shi2026,Aster2026}. A massive star origin has been ruled out for all three objects, and binary systems with lower-mass progenitors are preferred. A binary system has also been proposed for SN~2023fyq \citep{Dong2024}. These observations strongly indicate that the progenitor channels for SNe Ibn/Icn are likely diverse, with both massive single stars and stripped binaries contributing. 

The above diversity is also seen in the spectral properties of SNe Ibn. Two spectral classes have been proposed based on whether the \ion{He}{i} lines are seen only in emission or with P Cygni profiles \citep{Hosseinzadeh2017,Dong2025}. For the first class (e.g. SN~2006jc) the absence of broader features suggests a strong deceleration of the ejecta by the CSM due to a weak explosion and/or a dense and extended CSM \citep{Dessart2016,Dessart2022}, while for the second class the broader spectral features point towards an ejecta that have swept through most of the CSM and therefore there is a less extended and lower density CSM and/or a more energetic explosion. Flash features---which require dense, confined CSM at the time of shock breakout---are more common in the second group \citep[see Figure~10 in][]{Dong2025}. \SN{} belongs to the second class due its strong P~Cygni profiles and the appearance of ejecta features. Overall, there is no clear mapping between these spectroscopic classes \citep[and other observables, see][]{Dong2025} and the various progenitor channels for SNe Ibn, which likely reflects a continuum in the CSM properties applicable to both single-star and binary scenarios. 

\SN{} likely occurred within a region of active star formation given its late-type host galaxy and location (Section \ref{sec:host}). Our bolometric light curve modelling yielded a low \textsuperscript{56}Ni mass of ${\sim}0.02~M_\odot$, consistent with the SN Ibn sample in \citet{Maeda2022} and therefore potentially consistent with a massive progenitor ($M_\mathrm{ZAMS}{\gtrsim}18~M_\odot$) with substantial fallback onto a BH. The \ion{C}{iii} lines seen in our first spectrum---more commonly seen in SNe Icn---might point towards a progenitor closer to the hypothesised transition mass of $M_\mathrm{ZAMS}{\sim}40$--50$~M_\odot$. Furthermore, given that wave-driven mass loss in a high-mass progenitor (${\geq}30~M_\odot$, Section \ref{subsec:precursor} above) can explain the precursor timescale and energetics we suggest a progenitor in the range $M_\mathrm{ZAMS}{\sim}30$--40$~M_\odot$, although we cannot exclude a binary origin.   

Looking forward, the Vera Rubin Observatory's LSST will obtain light curves of pre-SN activity with an unprecedented level of detail. Although spectroscopy during these phases will be challenging, a precursor analogous to \SN{} would have been within the capabilities of 8 m-class telescopes. Observations during these phases will be critical for unveiling the physical mechanisms responsible for triggering extreme mass-loss in the final years to months of certain massive stars.

\section{Conclusions}
In this paper we presented observations of the Type Ibn SN~2025kzr at 51 Mpc. Our high cadence spectroscopic dataset enabled a detailed study of the flash-ionisation phase, allowing us to constrain the CSM properties and investigate its connection to the precursor outburst. Here we summarise our main findings:
\begin{enumerate}

\item Archival imaging shows a precursor outburst beginning ${\sim}55$~days before explosion with a peak brightness of $M_r\approx-13.6$~mag. We estimate a peak luminosity of ${\approx}10^{41}$~erg~s$^{-1}$ and a total radiated energy of ${\approx}2\times10^{47}$~erg over the duration of the precursor. \SN{} is only the fourth SN Ibn showing precursor emission after SNe 2006jc, 2019uo, and 2023fyq.

\item Photometrically, \SN{} showed fast-rising, UV-bright emission with light curves peaking earlier in bluer bands compared to redder bands (6.1 days in $uvw2$ versus 13.0 days in $z$). The peak (extinction-corrected) $r$-band brightness of $M_r=-19.26\pm0.09$~mag and light curve evolution do not stand out compared to previous SNe Ibn. The $g-r$ and $B-V$ colour curves show a flattening at ${\sim}20$ days post-peak suggestive of a new power-source taking over.   

\item The UV/optical photometric SEDs are well-approximated by blackbody emission throughout their evolution, reaching a peak temperature of $T{\approx}29000$~K at 3.7 days and a peak in the blackbody luminosity of ${\sim}10^{44}$~erg~s$^{-1}$ at 5.8 days. After peak, the photospheric temperature declines as $t^{-0.9}$ until 40 days. The post-peak bolometric light curve can be fitted with a broken power-law with an initial decline of $t^{-2.0}$ followed by a steepening to $t^{-5.2}$ . The break time of 25 days coincides with a peak in the radius evolution. 

\item The late-time ($t>40$~d) bolometric light curve shows a flattening that can be explained by radioactive power, and we infer an upper limit on the mass of synthesised \textsuperscript{56}Ni of ${<}0.06~M_\odot$ assuming a canonical explosion energy of $10^{51}$~erg with $2~M_\odot$ of ejecta. Fitting the ejecta-CSM interaction model of \citet{Jiang2020} to our bolometric light curve we find $M_\mathrm{ej}{\sim}1.6~M_\odot$, $M_\mathrm{CSM}{\sim}1.7~M_\odot$, and $M_\mathrm{Ni}{\sim}0.02~M_\odot$, although we note that this model assumes a constant CSM density rather than the steeper density profile ($\rho_\mathrm{CSM}\propto r^{-3}$) preferred by \citet{Maeda2022}. 

\item Our high-cadence spectroscopic dataset spanning 1.9--58.5 days shows a number of distinct phases. Flash-ionised emission features of \ion{He}{ii}, \ion{C}{iv}, \ion{N}{iv}, \ion{C}{iii}, and \ion{N}{iii} persist until 10 days. The higher-ionisation lines reach peak strength at 3.8 days, coincident with the peak photospheric temperature. Thereafter the spectra are dominated by prominent \ion{He}{i} P Cygni lines with absorption minima at a velocity of 1500~km~s$^{-1}$. By 23.8 days the spectrum shows broader ejecta-like features reminiscent of SNe Ic, albeit with narrow \ion{He}{i} emission lines superposed. The late-time spectra ($t>50$~days) resemble SN~2010al at a similar phase and are characterised by a prominent \ion{Ca}{ii} NIR triplet in emission along with broader \ion{He}{i} features with absorption minima at ${\sim}5000$~km~s$^{-1}$. 

\item Our early flash spectra resemble early-type WN Wolf-Rayet spectra, but with emission lines showing broader wings (FWHM$\approx4000$--6000~km~s$^{-1}$) due to electron scattering. Using the Pickering-decrement method we find a CSM composition that is fully H-free. Furthermore, our 3.8-day SALT spectrum clearly shows a blueshift of the \ion{He}{ii} line centres by 460~km~s$^{-1}$ compared to the \ion{He}{i} lines which are at zero velocity. This can be explained by the occultation of the far side of the CSM by the photosphere. The blueshift of the higher-ionisation lines appears to diminish as the flash lines weaken.

\item Polarimetric observations at 4.9 days---while the flash-ionisation phase was ongoing---are consistent with a spherically symmetric CSM.  At 10.9 days we observe significant polarisation ($P{\sim}0.74\%$), indicative of a deviation from spherical symmetry consistent with the emerging Ic-like spectrum seen later in the spectral evolution.  

\end{enumerate}
The timescale for the disappearance of the flash lines together with the observed CSM speed imply a pre-SN mass ejection occurring ${\approx}66$~days before explosion. Given that we detect a precursor outburst 55~days before explosion this strongly suggests that the CSM probed by our flash spectroscopy was ejected during the precursor itself. From our flash spectroscopy and photometry we estimate a CSM mass in the range $0.03$--$1.7~M_\odot$, implying a high mass-loss rate of ${\gtrsim}10^{-1}~M_\odot$~yr$^{-1}$ caused by an extreme mass-loss event. The timescale and energetics of the precursor can be explained by wave-driven mass loss related to the late stages of nuclear burning---in particular the oxygen-burning phase---which requires a progenitor with $M_\mathrm{ZAMS}\geq30~M_\odot$ in the single-star scenario. A high-mass progenitor is supported by the low \textsuperscript{56}Ni mass inferred from the bolometric light curve, the \ion{C}{iii} emission lines in our early spectra that are more commonly seen in SNe Icn, and the fully H-free composition for the confined CSM. Taken together, we favour a single massive WR progenitor with $M_\mathrm{ZAMS}{\sim}30$--$40~M_\odot$, although we cannot exclude a binary origin. Spectroscopic observations during pre-SN outbursts detected by the Vera Rubin LSST will be critical to unveiling the physical mechanisms behind such extreme mass loss events.  

\section*{Data availability}
The photometric data for \SN{} presented in Figure \ref{fig:LC_all}, as well as the bolometric light curve along with the temperature and radius evolution presented in Figure \ref{fig:Lbol} are available at the CDS via https://cdsarc.cds.unistra.fr/viz-bin/?. All spectra presented in Figure \ref{fig:all_spec} are publicly available at WISeREP.

\begin{acknowledgements}
GL and SdW were supported by a research grant (VIL60862) from VILLUM FONDEN. PC acknowledges the support from the Zhejiang provincial top-level research support program. MP acknowledges support from a UK Research and Innovation Fellowship (UKRI1062). TLK acknowledges support via a Warwick Astrophysics prize post-doctoral fellowship, made possible thanks to a generous philanthropic donation. CPG acknowledges financial support from grant RYC2024-050959-I, funded by MICIU/AEI/10.13039/501100011033 and the FSE+, as well as from projects PID2023-151307NB-I00, PIE 20215AT016, and CEX2020-001058-M, and the MaX-CSIC Excellence Award MaX4-SOMMA-ICE. SJS acknowledges funding from STFC Grants ST/Y001605/1, ST/X001253/1, a Royal Society Research Professorship and the Hintze Family Charitable Foundation. TEMB is funded by Horizon Europe ERC grant no. 101125877. KWS acknowledges funding from the Royal Society. TWC acknowledges financial support from the Yushan Fellow Program of the Ministry of Education, Taiwan (MOE-111-YSFMS-0008-001-P1), and from the National Science and Technology Council, Taiwan (NSTC 114-2112-M-008-021-MY3). MN is supported by the European Research Council (ERC) under the European Union's Horizon 2020 research and innovation programme (grant agreement No.~948381). 

This work is based on observations made at the South African Astronomical Observatory (SAAO) through the "Intelligent Observatory" rapid follow-up program. The SAAO is funded by the South African National Research Foundation (NRF). Some of the observations reported in this paper were obtained with the Southern African Large Telescope (SALT), under programme 2024-2-LSP-001 (PI: DAHB). Polish participation in SALT is funded by grant No. MNiSW DIR/WK/2016/07. The MeerKAT telescope is operated by the South African Radio Astronomy Observatory, which is a facility
of the National Research Foundation, an agency of the Department of Science and Innovation. This work has made use of the “MPIfR S-band receiver system” designed, constructed and maintained by funding of the MPI für Radioastronomy and the Max-Planck-Society. This work has made use of data from the Asteroid Terrestrial-impact Last Alert System (ATLAS) project. The Asteroid Terrestrial-impact Last Alert System (ATLAS) project is primarily funded to search for near earth asteroids through NASA grants NN12AR55G, 80NSSC18K0284, and 80NSSC18K1575; byproducts of the NEO search include images and catalogs from the survey area. This work was partially funded by Kepler/K2 grant J1944/80NSSC19K0112 and HST GO-15889, and STFC grants ST/T000198/1 and ST/S006109/1. The ATLAS science products have been made possible through the contributions of the University of Hawaii Institute for Astronomy, the Queen's University Belfast, the Space Telescope Science Institute, the South African Astronomical Observatory, and The Millennium Institute of Astrophysics (MAS), Chile. Pan-STARRS is primarily funded to search for near-Earth asteroids through NASA grants NNX08AR22G and NNX14AM74G. The Pan-STARRS science products were made possible through the contributions of the University of Hawai'i Institute for Astronomy, the Queen's University Belfast and the University of Oxford. This work made use of data supplied by the UK Swift Science Data Centre at the University of Leicester. The ZTF forced-photometry service was funded under the Heising-Simons Foundation grant \#12540303 (PI: Graham). Based on observations made with ESO Telescopes at the La Silla Observatory under programme ID 112.25JQ.018, 112.25JQ.019, 115.27YA.001. Based on observations made with the Nordic Optical Telescope, owned in collaboration by the University of Turku and Aarhus University, and operated jointly by Aarhus University, the University of Turku and the University of Oslo, representing Denmark, Finland and Norway, the University of Iceland and Stockholm University at the Observatorio del Roque de los Muchachos, La Palma, Spain, of the Instituto de Astrofisica de Canarias. The NOT data were obtained under program ID P71-021. The data presented here were obtained [in part] with ALFOSC, which is provided by the Instituto de Astrofisica de Andalucia (IAA) under a joint agreement with the University of Copenhagen and NOT.

\end{acknowledgements}
\bibliographystyle{aa}
\bibliography{aas}

\begin{appendix}

\section{Table of spectroscopic observations}
\begin{table*}
\caption{Log of spectroscopic observations}
\label{tab:obs_log}
\centering
\begin{tabular}{lllllll}
\hline\hline
Date (UTC)  & MJD & Phase\tablefootmark{a} (d) & Telescope/Instrument & Observing setup & Exp. time (s)  \\
\hline
2025-05-22 20:24:40 & 60817.85 & 1.85 & Lesedi/Mookodi & mk-wide & 600 \\
2025-05-23 19:38:13 & 60818.82 & 2.82 & Lesedi/Mookodi & mk-narrow & 2400 \\
2025-05-24 18:42:57 & 60819.78 & 3.78 & Lesedi/Mookodi & mk-narrow & 2400 \\
2025-05-24 19:26:54 & 60819.82 & 3.82 & SALT/RSS & PG0900/$13.25^{\circ}$ & 1200  \\
2025-05-24 19:47:44 &  &  &  & PG0900/$20.00^{\circ}$ & 1200\\
2025-05-24 23:27:37 & 60820.00 & 4.00 & NTT/EFOSC2 & Gr\#11 & 1500 \\
2025-05-24 23:53:26 &  &  &   & Gr\#16 & 1500 \\
2025-05-25 19:08:48 & 60820.80 & 4.80 & Lesedi/Mookodi & mk-wide & 600 \\ 
2025-05-25 21:47:05 & 60820.91 & 4.91 & NOT/ALFOSC & Gr\#4 & 1500 \\
2025-05-25 23:27:49 & 60820.99 & 4.99 & NTT/EFOSC2 & Gr\#11 & 1200 \\
2025-05-25 23:48:38 &  &  & & Gr\#16 & 1200 \\
2025-05-26 18:43:39 & 60821.78 & 5.78 & Lesedi/Mookodi & mk-wide & 1200 \\ 
2025-05-26 19:35:21 & 60821.82 & 5.82 & SALT/HRS & Low-res & 1800 \\
2025-05-26 21:59:49 & 60821.92 & 5.92 & NOT/ALFOSC & Gr\#4 & 1500 \\
2025-05-27 16:41:34 & 60822.70 & 6.70 & Lesedi/Mookodi & mk-wide & 600 \\
2025-05-28 17:55:26 & 60823.75 & 7.75 & Lesedi/Mookodi & mk-wide & 600 \\
2025-05-30 01:33:41 & 60825.08 & 9.08 & NTT/EFOSC2 & Gr\#11 & 900 \\
2025-05-30 01:49:31 &  &  & & Gr\#16 & 900 \\
2025-05-30 16:54:19 & 60825.70 & 9.70 & Lesedi/Mookodi & mk-wide & 1200 \\
2025-05-30 23:40:49 & 60826.00 & 10.00 & NTT/EFOSC2 & Gr\#11 & 900 \\
2025-05-30 23:56:38 &  &  & & Gr\#16 & 900 \\
2025-05-31 16:57:03 & 60826.71 & 10.71 & Lesedi/Mookodi & mk-wide & 600 \\
2025-06-01 17:51:05 & 60827.74 & 11.74 & Lesedi/Mookodi & mk-wide & 600 \\
2025-06-01 20:58:01 & 60827.87 & 11.87 & NOT/ALFOSC & Gr\#4 & 1200 \\
2025-06-02 16:43:59 & 60828.70 & 12.70 & Lesedi/Mookodi & mk-wide & 600 \\
2025-06-03 17:10:57 & 60829.72 & 13.72 & Lesedi/Mookodi & mk-wide & 600 \\
2025-06-04 17:38:28 & 60830.74 & 14.74 & Lesedi/Mookodi & mk-wide & 600 \\
2025-06-05 18:47:13 & 60831.80 & 15.80 & SALT/RSS & PG0900/$13.25^{\circ}$ & 1200 \\
2025-06-05 19:08:02 &  &  &  & PG0900/$20.00^{\circ}$ & 1140 \\
2025-06-05 20:18:49 & 60831.85 & 15.85 & Lesedi/Mookodi & mk-wide & 600 \\
2025-06-07 21:14:48 & 60833.89 & 17.89 & NOT/ALFOSC & Gr\#4 & 1200 \\
2025-06-10 16:51:46 & 60836.70 & 20.70 & Lesedi/Mookodi & mk-wide & 900 \\
2025-06-10 17:56:27 & 60836.75 & 20.75 & SAAO 1.9m/SpupNIC  & Gr\#6 & 1800 \\
2025-06-11 17:17:44 & 60837.72 & 21.72 & Lesedi/Mookodi & mk-wide & 900 \\
2025-06-12 16:56:16 & 60838.71 & 22.71 & Lesedi/Mookodi & mk-wide & 900 \\
2025-06-13 18:13:46 & 60839.77 & 23.77 & SALT/RSS & PG0900/$13.25^{\circ}$ & 900 \\
2025-06-13 18:32:25 &  &  &  & PG0900/$20.00^{\circ}$ & 900 \\
2025-06-13 19:17:17 & 60839.80 & 23.80 & Lesedi/Mookodi & mk-wide & 900 \\
2025-06-14 18:45:28 & 60840.78 & 24.78 & Lesedi/Mookodi & mk-wide & 900 \\
2025-06-15 16:39:31 & 60841.69 & 25.69 & Lesedi/Mookodi & mk-wide & 900 \\
2025-06-16 18:13:54 & 60842.76 & 26.76 & Lesedi/Mookodi & mk-wide & 900 \\
2025-06-19 18:23:17 & 60845.77 & 29.77 & Lesedi/Mookodi & mk-wide & 900 \\
2025-06-21 18:55:43 & 60847.79 & 31.79 & Lesedi/Mookodi & mk-wide & 900 \\
2025-06-22 17:58:52 & 60848.75 & 32.75 & Lesedi/Mookodi & mk-wide & 900 \\
2025-07-12 23:03:58 & 60868.96 & 52.96 & NTT/EFOSC2 & Gr\#11 & 3600 \\
2025-07-16 00:30:44 & 60872.02 & 56.02 & NTT/EFOSC2 & Gr\#16 & 1553 \\
2025-07-18 11:45:36 & 60874.49 & 58.49 & Magellan/IMACS & Grism-300/$17.5^{\circ}$ & 900\\
\hline
\end{tabular}
\tablefoot{
\tablefoottext{a}{Phase with respect to MJD 60816.0.}
}
\end{table*}
\end{appendix}

\end{document}